\documentclass[10pt,a4]{amsart}
\usepackage{times}
\normalfont
\usepackage[T1]{fontenc}
\usepackage{graphics}
\usepackage{graphicx}
\usepackage{amsmath, amssymb,url}
\begin{document}
\title{An introduction to spectral distances in networks (extended version)}

\author{Giuseppe Jurman}
\address{Fondazione Bruno Kessler, Trento, Italy}
\email{jurman@fbk.eu}

\author{Roberto Visintainer}
\address{Fondazione Bruno Kessler and DISI University of Trento, Trento, Italy}
\email{visintainer@fbk.eu}

\author{Cesare Furlanello}
\address{Fondazione Bruno Kessler, Trento, Italy}
\email{furlan@fbk.eu}

\begin{abstract}
Many functions have been recently defined to assess the similarity among networks as tools for quantitative comparison. They stem from very different frameworks - and they are tuned for dealing with different situations. Here we show an overview of the spectral distances, highlighting their behavior in some basic cases of static and dynamic synthetic and real networks.
\end{abstract}
\maketitle

\section*{Introduction}

Citing a comprehensive review \cite{boccaletti06complex}, a complex
network is a graph whose structure is irregular and dynamically
evolving in time. In terms of architectures, Strogatz
\cite{strogatz01exploring} used the term ''complex'' to describe a
network that is the counterpart of ''regular'' graphs (chains, grids,
lattices and fully-connected graphs), the random graphs lying at the
extremal edge of the complexity spectrum. Network models from
empirical studies lie somewhere in between regularity and randomness;
although more often unbalanced towards the latter, they can have to
unexpectedly highly symmetric structures \cite{macarthur08symmetry}.

This article reviews and benchmarks a class of methods that tackle the
problem of comparing structure between networks. Structure and
structural properties of networks have been studied in a wide variety
of fields in science \cite{albert02statistical, newman03structure,
  liben-nowell05algorithmic, boccaletti06complex}, with methods
ranging from statistical physics to machine learning
\cite{dorogovtsev08critical,goldenberg09survey}.  Structural analysis
is of central importance in computational biology
\cite{lacroix08introduction}.  Cootes pointed out that the comparison
of biological networks can provide much more evolutionary information
than studying each network separately
\cite{cootes07identification}. Furthermore, the comparison of protein
interaction networks can help designing models of cellular functions
\cite{vazquez03global,sharan06modeling}. Comparison methods are
essential with dynamic networks to measure differences between two
consecutive network states and then model the whole series. Comparison
is also essential in network reconstruction (e.g. of gene regulation
networks) by structure reverse engineering starting from steady-state
or time series data \cite{he09reverse, marbach10revealing,
  lee09computational}, where performance has to be gauged against the
ground truth of a real or simulated network.

Our interest for network comparison is motivated by the study of
network stability. On this less beaten path, only network robustness
with respect to perturbations has been considered until now
\cite{tercieux06characterization,pomerance09effect}. It is envisioned
that the choice of appropriate measures between networks would enable
new model selection procedures such those available in molecular
profiling for sets of ranked gene lists \cite{jurman08algebraic}.

In this study, six candidate distances derived from the family of
spectral similarity measures are investigated for network
comparison. After a first presentation of spectral measures and
alternatives in the rest of this introduction, a technical overview is
provided in Sect. \ref{sec:overview} and candidate measures are
presented. Benchmark data and experiments devised to exemplify and
compare the candidates are presented in Sect. \ref{sec:results}. 

\subsection*{Related Work}

The basic goal of network comparison is quantifying difference between
two homogeneous objects in some network space.  The theory of network
measurements relies on the quantitative description of main properties
such as degree distribution and correlation, path lenghts, diameter,
clustering, presence of motives \cite{costa07characterization}. These
and other properties have been described for complex networks in
\cite{newman03structure, boccaletti06complex} and recently reviewed by
MacArthur and S\'{a}nchez-Garc\'\i{}a \cite{macarthur09spectral}. Furthermore, network measurements can
be encoded into a feature vector, yelding a representation convenient
for classification tasks \cite{costa07what}.
\par
The use of similarity measures on the topology of the underlying
graphs defines a different strategy, whose roots date back to the 70's
with the theory of graph distances (regarding both metrics inter- and
intra-graphs \cite{entringer76distance}). Since then, a number of
similarity measures have been introduced, including metrics relaxed to
less stringent bounds. Cost-based functions stems from the parallel
theory of graph alignment: the edit distance and its variants use
the minimum cost of transformation of one graph into another by
means of the usual edit operations - insertion and deletion of links.
\par
Feature-based measures are instead obtained when the similarity
function is based on measurements feature vectors. One notable example
in this family is the recently proposed use of $\zeta$-functions for
network volume measurements \cite{shanker07defining, shanker07graph}.
\par
Finally, the label ``structure-based'' distance groups all other
measures that do not rely on cost functions or characteristic
features.  A typical example are those measures based on functions of
the maximal common subgraphs between the two networks, or those based
on the common motifs \cite{milo02network}, i.e.  patterns of
interconnections occurring in complex networks significantly more
often than in randomized networks. Remarkably, equivalence of some
structure-based distance and the edit distance has been proven
\cite{bunke97relation}). Although in most cases only network topology
is considered, measures were also introduced that deal
with directed or weighted links: for an example of a generic
construction and an application to biological networks, see
\cite{ahnert08applying}.

The family of spectral measures, which is investigated in this paper,
is also part of the group of structure-based distances. Basically, it
consists of a variety of maps of network's eigenvalues. The theory of
graph spectra started in the early 50's and since then many of its
aspects have been deeply mined, including a first classification of
networks \cite{banerjee08spectral}. The spectral theory has been
applied to biological networks \cite{banerjee07spectral,
  banerjee09graph}, where the properties of being scale-free (the
degree distribution following a power law) and small-world (most nodes
are not neighbors of one another, but most nodes can be reached from
every other by a small number of hops or steps) are particularly
evident. Estimates (also asymptotic) of the eigenvalues distribution
are available for complex networks \cite{rodgers05eigenvalue}.  The
idea of using spectral measures for network comparison is instead only
recent and it relies on similarity measures that are functions of the
network eigenvalues. However, it is important to note that, because of the existence of
isospectral networks, all these measures are indeed distances between classes
of iosospectral graphs.  An overview of the most common spectral
similarity measures and of their basic properties is presented in the rest of this paper. 

\section{Notations}
\label{sec:notations}
\begin{figure}[t]
\begin{tabular}{cccc}
\includegraphics[width=2.5cm]{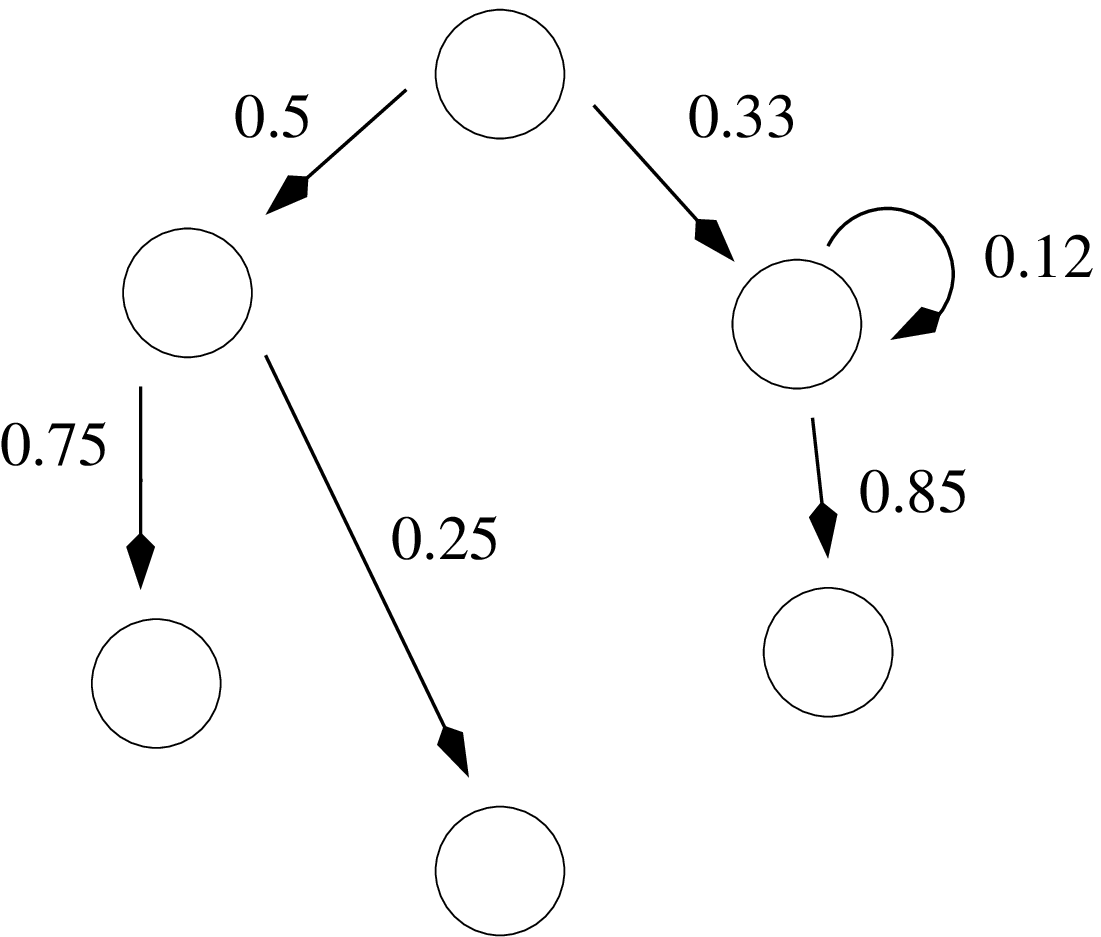}&
\includegraphics[width=2cm]{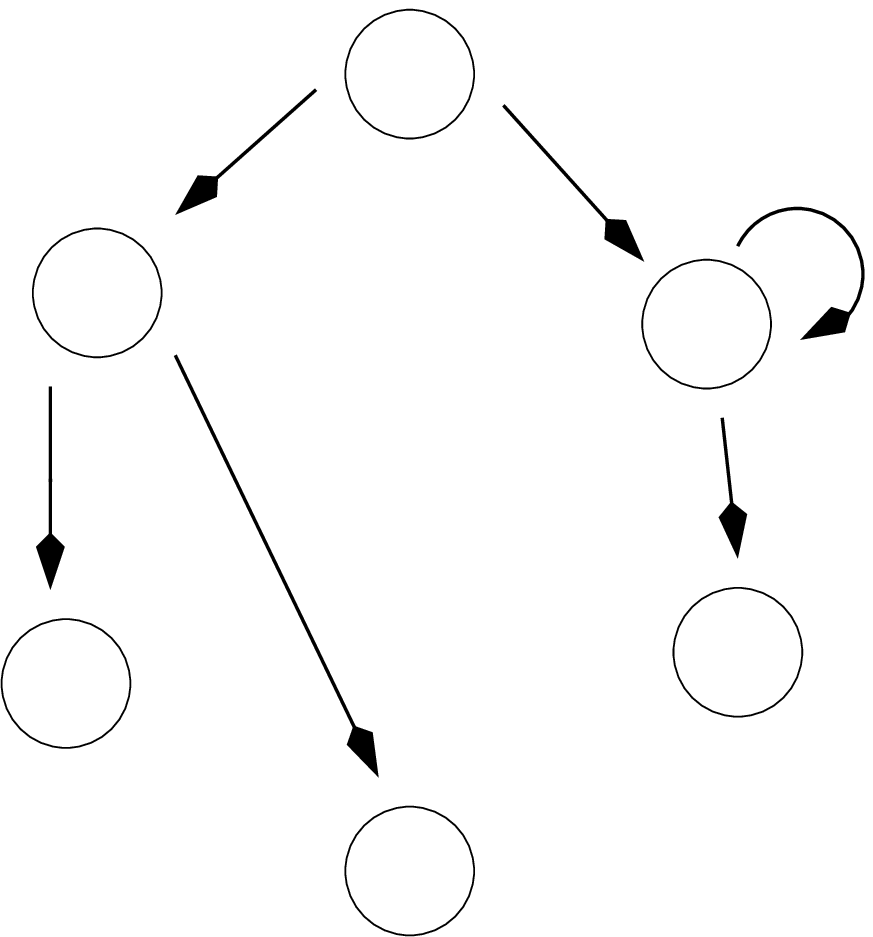}&
\includegraphics[width=2.5cm]{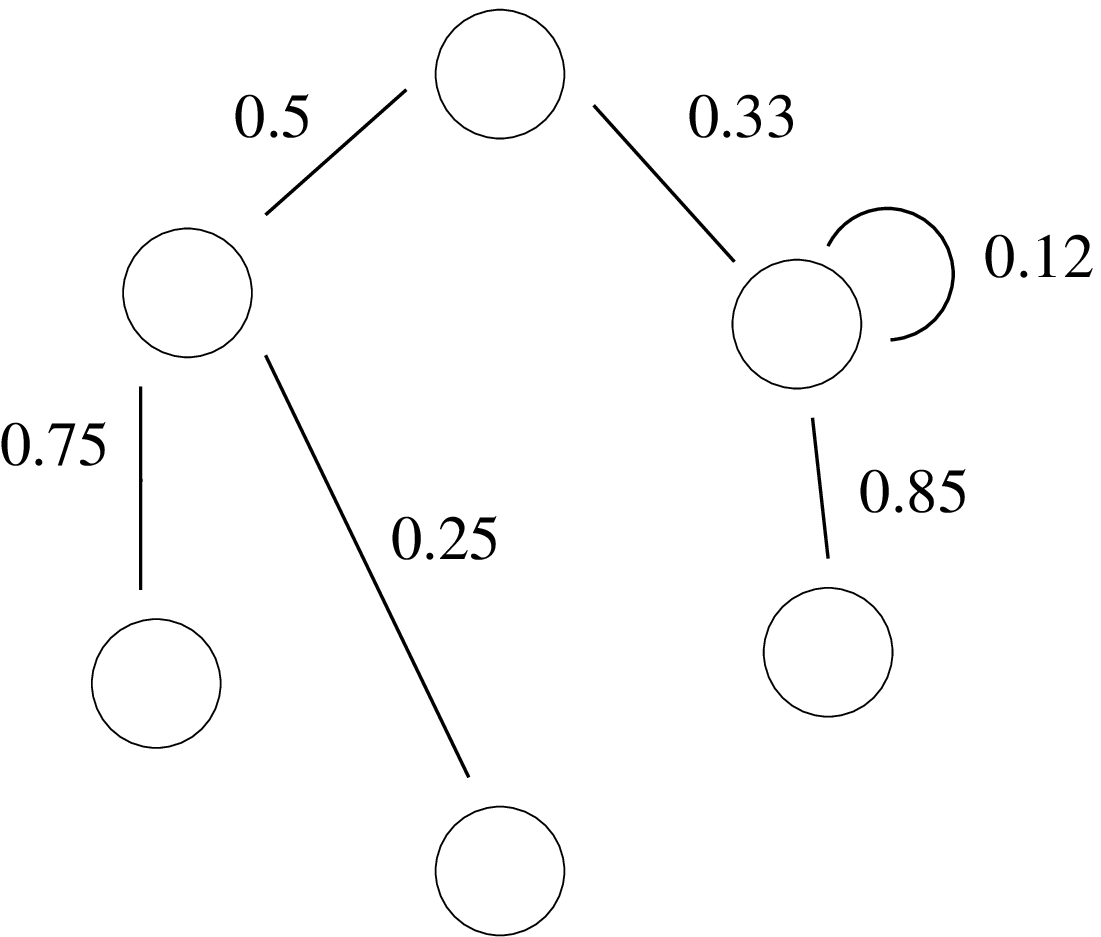}&
\includegraphics[width=2cm]{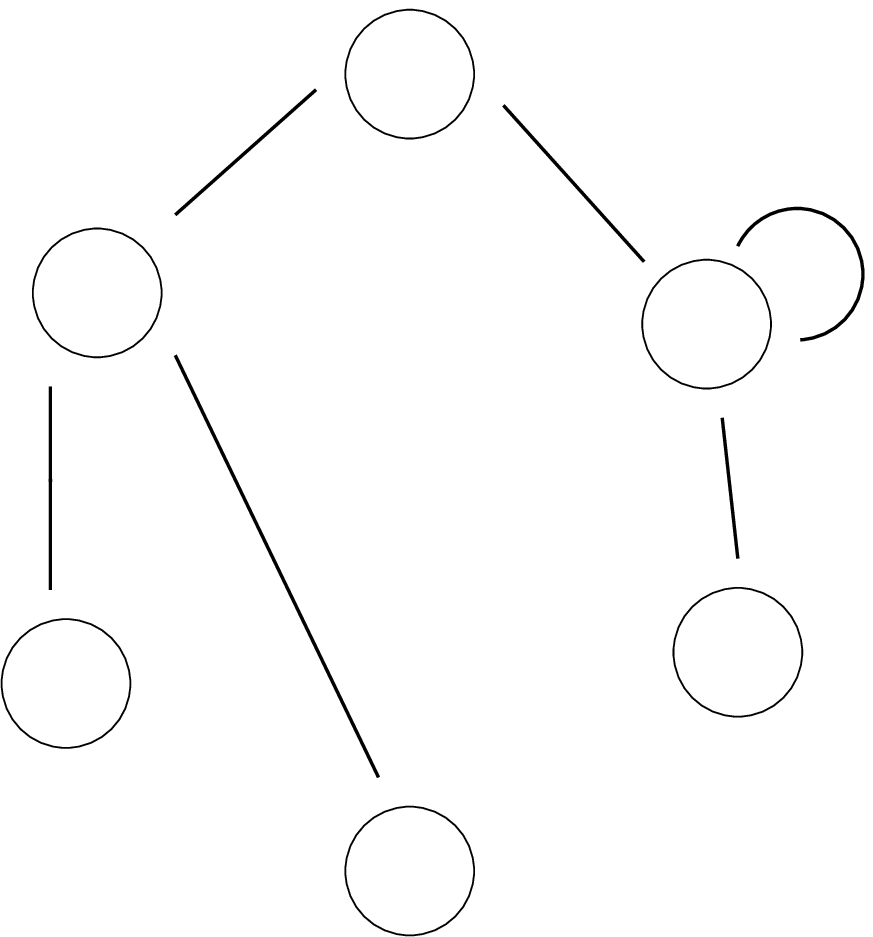}\\
Weighted digraph & Unweighted digraph & Weighted graph & Underlying topology
\end{tabular}
\caption{Network types}
\label{fig:networks}
\end{figure}

Formally, any network can be represented as a graph, a mathematical entity consisting of $N$ nodes (vertices) and $E$ edges (links or arrows) connecting pairs of nodes and representing interactions ($N\in\mathbb{N}\cup\{\infty\}$). 
Loops are allowed, i.e. an edge can link the same node to indicate self-interaction (some authors use the term pseudograph to indicate graph with loops).
Edges can be bidirectional or unidirectional: in the latter case the graph is called directed (digraph, for short) and the edges are represented by arrows.
Moreover, edges can carry weights to indicate interaction intensity: in this case, the network is called weighted.
More refined structures exist but they are not considered here. For instance: labeled graphs, where functions from some subsets of the integers to the vertices (edges) of the graph identify classes of vertices (edges); hypergraphs, where an edge can connect any number of vertices; and multigraphs, where any numbers of edges between two vertices are allowed.
%% Even more refined structures can be considered: for instance labeled graphs, where functions from some subset of the integers to the vertices (edges) of the graph identify classes of vertices (edges) hypergraphs, where an edge can connect any number of vertices (not just two), but they are not considered here and multigraphs, where any numbers of edges between two vertices are allowed.
For any network $G$, its topology consists of the set $V(G)=\{v_1,\dots,v_N\}$ of its nodes and the set $E(G)=\{e_1=(v_{i_1},v_{j_1}), \cdots e_E=(v_{i_E},v_{j_E})\}$ of its edges, neglecting weights and directions.
Different types of graph sharing the same topology are displayed in Fig. \ref{fig:networks}.

A network, or graph, is characterized completely by its adjacency matrix $A$, i.e. an $N\times N$ matrix whose nonzero entries denote the various links between the graph's $N$ nodes.
Directions and weights are represented by the signs (or by asymmetricity) and values of the matrix entries.
For the underlying topology (and thus for any unweighted undirected network), the adjacency matrix is symmetric and with entries in $\{0,1\}$.
The adjacency matrices for the weighted digraph in Fig. \ref{fig:networks} and its topology are shown in Tab. \ref{tab:adjacency}, where nodes ordering is clockwise starting from the top node.
This representation is not unique, in that it depends on the actual labeling of the nodes, and isomorphic graphs (identical graphs with permuted labels) share the same adjacency matrix.
Similarly, graphical representations are not unique too, since node placement is arbitrary.

\begin{table}[b]
\caption{Adjacency matrices for the weighted directed network (two alternative matrices, with sign indicating direction or asymmetric, with the (positive) value only in entry $(i,j)$ if $i\to j$) in Fig. \ref{fig:networks} and its topology; nodes ordering is clockwise starting from the top node.}
\label{tab:adjacency}
\begin{center}
\begin{tabular*}{0.85\textwidth}{cc}
\hline
\textbf{Network} & \textbf{Adjacency matrix}\\
\hline
\begin{tabular}{c}
\includegraphics[width=2.5cm]{./figures/top-ori-wei.eps}
\end{tabular}
&
$\begin{pmatrix}
0 & 0.33 & 0 & 0 & 0& 0.5 \\
(-0.33) & 0.12 & 0.85 & 0 & 0 & 0 \\
0 & (-0.85) & 0 & 0& 0& 0 \\
0& 0&0&0&0& (-0.25) \\
0& 0&0&0&0& (-0.75) \\
(-0.5) & 0 &0 & 0.25 & 0.75 &0
\end{pmatrix}
$
\\
\hline
\begin{tabular}{c}
\includegraphics[width=2cm]{./figures/top.eps}
\end{tabular}
&
$
\begin{pmatrix}
\phantom{-}0\phantom{.00} & \phantom{-}1\phantom{.00} & \phantom{-}0\phantom{.00} & \phantom{-}0\phantom{.00} & \phantom{-}0\phantom{.00} & \phantom{-}1\phantom{.00} \\
\phantom{-}1\phantom{.00} & \phantom{-}1\phantom{.00} & \phantom{-}1\phantom{.00} & \phantom{-}0\phantom{.00} & \phantom{-}0\phantom{.00} & \phantom{-}0\phantom{.00} \\
\phantom{-}0\phantom{.00} & \phantom{-}1\phantom{.00} & \phantom{-}0\phantom{.00} & \phantom{-}0\phantom{.00} & \phantom{-}0\phantom{.00} & \phantom{-}0\phantom{.00} \\
\phantom{-}0\phantom{.00} & \phantom{-}0\phantom{.00} & \phantom{-}0\phantom{.00} & \phantom{-}0\phantom{.00} & \phantom{-}0\phantom{.00} & \phantom{-}1\phantom{.00} \\
\phantom{-}0\phantom{.00} & \phantom{-}0\phantom{.00} & \phantom{-}0\phantom{.00} & \phantom{-}0\phantom{.00} & \phantom{-}0\phantom{.00} & \phantom{-}1\phantom{.00} \\
\phantom{-}1\phantom{.00} & \phantom{-}0\phantom{.00} & \phantom{-}0\phantom{.00} & \phantom{-}1\phantom{.00} & \phantom{-}1\phantom{.00} & \phantom{-}0\phantom{.00} \\
\end{pmatrix}
$
\\
\hline
\end{tabular*}
\end{center}
\end{table}

The degree ($\textrm{deg}$) of a vertex in an undirected graph is the number of edges touching the vertex itself, with loops (usually, but not for all authors) counted twice. The degree matrix is the diagonal matrix with the vertex degrees: for instance, for the network topology in Fig. \ref{fig:networks}, the degree matrix is $D = \left(\begin{smallmatrix} 2\\ &4\\ &&1\\ &&&1\\ &&&&1\\ &&&&&3\end{smallmatrix}\right)$. 
The Laplacian matrix of a graph is defined as the difference between the degree and the adjacency matrices: $L=D-A$. 
Thus, for an undirected and unweighted graph with no loops (a simple graphs), $L$ has zero row/column sum.

There exist at least two different normalized versions of the Laplacian matrix, namely $\mathcal{L}=D^{-\frac{1}{2}}LD^{-\frac{1}{2}}=I-D^{-\frac{1}{2}}AD^{-\frac{1}{2}}$ and $\Delta=D^{\frac{1}{2}}\mathcal{L} D^{-\frac{1}{2}}$, where $I$ is the identity matrix and $D^{-\frac{1}{2}}$ is the diagonal matrix with entries $-\frac{\delta_{ij}}{\sqrt{\textrm{deg}_i}}$.
In terms of the degree, their entries can be explicitely written as:
\begin{displaymath}
\mathcal{L}=
\begin{cases}
1 & \textrm{if $i=j$ and $\textrm{deg}_i\not =0$}\\
-\frac{1}{\sqrt{\textrm{deg}_i \textrm{deg}_j}} & \textrm{if $ij$ is an edge}\\
0 & \textrm{otherwise}
\end{cases}\;
\Delta=
\begin{cases}
1 & \textrm{if $i=j$ and $\textrm{deg}_i\not =0$}\\
-\frac{1}{\textrm{deg}_j} & \textrm{if $ij$ is an edge}\\
0 & \textrm{otherwise}
\end{cases}
\end{displaymath}
The matrices $\mathcal{L}$ and $\Delta$ are similar so they have the same set of eigenvalues (spectrum). 

The matrices $A, L, \mathcal{L}$ and $\Delta$ are called connectivity matrices of the graph.
An approach to connectivity matrices also in terms of the normalized Laplacian operators can be found in \cite{jost07dynamical, banerjee08spectrum,banerjee08spectrum2}.

An undirected and unweighted graph has symmetric real connectivity matrices and therefore real eigenvalues and a complete set of orthonormal eigenvectors. 
Also, for each eigenvalue, its algebraic multiplicity coincides with its geometric multiplicity.
Since A has zero diagonal, its trace and hence the sum of the eigenvalues is zero.
Moreover, $L$ is positive semidefinite and singular, so the eigenvalues are $0 = \mu_0 \leq \mu_1\leq \cdots\leq \mu_{n-1}$ and their sum (the trace of $L$) is twice the number of edges.
Finally, the eigenvalues of $\mathcal{L}$ lie in the range $[0,2]$.

While the connectivity matrices depend on the vertex labeling, the spectrum is a graph invariant.
Two graphs are called isospectral or cospectral if the corresponding connectivity matrices of the graphs have equal multisets of eigenvalues.
Isospectral graphs need not be isomorphic, but isomorphic graphs are always isospectral.
Network classification in terms of their spectrum is still an open problem \cite{vandam03which, wang06sufficient, wang10asymptotic}: however, a first attempt to (qualitative) network classification in terms of graph spectra can be found in \cite{banerjee08spectral, banerjee09spectral} by Banerjee.

For an introduction to the theory of graph spectra, see \cite{chung97spectral,cvetkovic10introduction, brouwer10spectra}. The relation between the spectral properties of the connectivity matrices and the structure and the dynamics of the networks are discussed in \cite{jost02evolving, jost07dynamical, almendral07dynamical}.

\section{Overview of spectral similarity measures}
\label{sec:overview}
In this section, we introduce a set of similarity measures based on the graph spectra that was recently proposed in literature, following an ideal chronological timeline.

The first distance D1 (or, indeed, one-parameter family of distances) we are presenting is possibly the most natural one. Originally D1 was introduced as an intra-graph measure \cite{jakobson02extremal, bacle05extremal} and mentioned as an inter-graph distance by Pincombe \cite{pincombe07detecting}, for evaluating changes in time-series of graphs.
Let $G$, $H$ be two graphs with $N$ nodes and let $\{\lambda_0=0\leq \lambda_1\leq\cdots\leq\lambda_{N-1}\}$, $\{\mu=0\leq \mu\leq\cdots\leq\mu_{N-1}\}$ the respective Laplacian spectra.
For an integer $k\leq N$, the distance is defined as:
\begin{equation}
\label{eq:pincombe07detecting}
d_k(G,H)=
\begin{cases}
{\sqrt{\frac{\displaystyle{\sum_{i=N-k}^{N-1}} (\lambda_i-\mu_i)^2}{\displaystyle{\sum_{i=N-k}^{N-1}} \lambda^2_i}}} & \textrm{if $\displaystyle{\sum_{i=N-k}^{N-1} \lambda^2_i\leq \sum_{i=N-k}^{N-1} \mu^2_i}$}\\
\\
{\sqrt{\frac{\displaystyle{\sum_{i=N-k}^{N-1}} (\lambda_i-\mu_i)^2}{\displaystyle{\sum_{i=N-k}^{N-1}} \mu^2_i}}}    & \textrm{if $\displaystyle{\sum_{i=N-k}^{N-1} \mu^2_i< \sum_{i=N-k}^{N-1} \lambda^2_i}$}\\
\end{cases}
\end{equation}
The D1 measure is non-negative, separated, symmetric and it satisfies the triangle inequality, so it is a measure.

A more refined spectral distance was defined as a step towards reconstructing a graph from its spectrum through a Metropolis algorithm \cite{ipsen02evolutionary}. 
The definition of the measure D2 follows the dynamical interpretation of a $N$-nodes network as a $N$-atoms molecules connected by identical elastic strings, where the pattern of connections is defined by the adjacency matrix of the corresponding network.
The dynamical system is described by the set of $N$ differential equations 
\begin{displaymath}
\ddot{x}_i+\sum_{j=1}^N A_{ij}(x_i-x_j)=0\quad\textrm{for\;$i=0,\cdots,N-1$}\ .
\end{displaymath}
The vibrational frequencies $\omega_i$ are given by the eigenvalues of the Laplacian matrix\footnote{In \cite{chung97spectral}, the Laplacian spectrum is called the vibrational spectrum.} of the network: $\lambda_i = \omega^2_i$, with $\lambda_0=\omega_0=0$.
The spectral density for a graph as the sum of Lorentz distributions is defined as
\begin{displaymath}
\rho(\omega)=K\sum_{i=1}^{N-1} \frac{\gamma}{(\omega-\omega_k)^2+\gamma^2}
\end{displaymath}
where $\gamma$ is the common width\footnote{The scale parameter $\gamma$ which specifies the half-width at half-maximum (HWHM), equal to half the interquartile range.} and $K$ is the normalization constant solution of
\begin{displaymath}
\int_0^\infty \rho(\omega)\textrm{d}\omega =1\ .
\end{displaymath}
Then the spectral distance $\epsilon$ between two graphs $G$ and $H$ with densities $\rho_G(\omega)$ and $\rho_H(\omega)$ can then be defined as 
\begin{equation}
\label{eq:ipsen02evolutionary}
\epsilon(G,H) = \sqrt{\int_0^\infty \left[\rho_G(\omega)-\rho_H(\omega)\right]^2} \textrm{d}\omega\ .
\end{equation}
Note that two above integrals can be explicitely computed through the relation $\displaystyle{\int \frac{1}{1+x^2}\textrm{d}x = \textrm{arctan}(x)}$.

A simpler measure D3 was introduced in \cite{zhu05study} for graph matching, using the graph edit distance as the reference baseline.
The authors compute the spectrum associated to the classical adjacency matrix, Laplacian matrix, signless Laplacian matrix $|L|=D+A$, and normalized Laplacian ($\mathcal{L}$) matrix. They also introduce two more functions: the path length distribution and the heat kernel $h_t$. 
The heat kernel is related to the Laplacian by the equation
\begin{displaymath}
\frac{\partial h_t}{\partial t} = -L h_t\ ,
\end{displaymath}
so that
\begin{displaymath}
h_t(u,v) = \sum_{i=0}^{N-1} e^{-\lambda_i t} \phi_i(u)\phi_i(v)\ ,
\end{displaymath}
where $\lambda_i$ are the Laplacian eigenvalues and $\phi_i$ the corresponding eigenvectors. 
For $t\to 0$, $h_t\to I-L t$, while when $t\to \infty$ then $h_t\to e^{-\lambda_{N-1} t}\phi_{N-1}{}^T\phi_{N-1}$. 
By varying $t$ different representation con be obtained, from the local ($t\to 0$) to the global ($t\to \infty$) structure of the network.
Moreover, if $D_k(u,v)$ is the number of paths of length $k$ between nodes $u$ and $v$, the following identity holds:
\begin{displaymath}
h_t(u,v) = e^{-t}\sum_{i=0}^{N^2-1} D_k(u,v)\frac{t^k}{k!}\ ,
\end{displaymath}
which allows the explicit computation of the path length distribution:
\begin{displaymath}
D_k(u,v) = \sum_{i=0}^{N-1} (1-\lambda_i)^k \phi_i(u)\phi_i(v)\ .
\end{displaymath}
The proposed distance is just the Euclidean distance between the vectors of (ordered) eigenvalues (for a given matrix $M$) for the two networks being compared:
\begin{equation}
\label{eq:zhu05study}
d_M(G,H) = \sqrt{\sum_{i=0}^{N-1} \left(\lambda^{(G,M)}_i -\lambda^{(H,M)}_i\right)^2}\ ,
\end{equation}
where $\lambda_{(T,M)}$ are the eigenvalues of the graph $T$ w.r.t. the matrix $M$, where $M$ is either a connectivity matrix, or the heat kernel matrix or the path length matrix.
As a final observation, the authors claim that the heat kernel matrix has the highest correlation with the edit distance, while the adjacency matrix hass the lowest.

A similar formula D4 is proposed in \cite{comellas08spectral} as the squared Euclidean ($L_2$) between the vectors of the Laplacian matrix:
\begin{equation}
\label{eq:comellas08spectral}
d(G,H) = \sum_{i=0}^{N-1}\left(\lambda^{(G,L)}_i -\lambda^{(H,L)}_i\right)^2\ .
\end{equation}

The next and last two measures are based on the concept of spectral distribution.

The distance D5 is introduced in \cite{fay09weighted}, aiming at comparing Internet networks topologies. 
Let $f_\lambda$ be the (normalized Laplacian) eigenvalued distribution, and $\mu(\lambda)$ a weighting function and define a generic distance between graphs $G$ and $H$ as follows
\begin{displaymath}
d_{\mu,p}(G,H) = \int_\lambda \mu(\lambda)\left(f_{\lambda,G}(\lambda)-f_{\lambda,H}(\lambda)\right)^p\textrm{d}{\lambda}\ .
\end{displaymath}
The weighting function is then defined as $\mu(\lambda)=(1-\lambda)^4$, an approximation of the graph irregularity as defined in \cite{chung97spectral}, while the usual Euclidean metric is chosen, so that $p=2$: the exact formula thus reads
\begin{equation}
\label{eq:fay09weighted_exact}
d(G,H) = \int_\lambda (1-\lambda)^4 \left(f_{\lambda,G}(\lambda)-f_{\lambda,H}(\lambda)\right)^2\textrm{d}{\lambda}\ .
\end{equation}
Calculating the eigenvalues of a large (even sparse) matrix is computationally expensive; an approximated version is also proposed, based on estimation of the distribution $f$ of eigenvalues by means of pivoting and Sylvester's Law of Inertia, used to compute the number of eigenvalues that fall in a given interval. 
To estimate the distribution $K$ equally spaced bins in the range $\left[ 0,2 \right]$ are used, so that a weighted spectral distribution measure for a graph $G$ can be defined for an integer $n>0$ as follows:
\begin{displaymath}
\omega_n(G) = \sum_{k\in K} (1-k)^n f(\lambda=k)\ .
\end{displaymath}
The generic formula can be now specialized to:
\begin{equation}
\label{eq:fay09weighted_approx}
d_n(G,H) = \sum_{k\in K} (1-k)^n (f_G(\lambda=k)- f_H(\lambda=k))^2\ ,
\end{equation}
a family of metrics parameterized by the integer $N$.

\begin{table}[t!]
\caption{Spectral graph distances}
\label{tab:distances}
\begin{center}
\begin{tabular}{cccc}
\hline
\textbf{Distance} & \textbf{Formula} & \textbf{Equation} & \textbf{Ref.} \\
\hline
\\
D1 & $d_k(G,H)=
\begin{cases}
{\sqrt{\frac{\displaystyle{\sum_{i=N-k}^{N-1}} (\lambda_i-\mu_i)^2}{\displaystyle{\sum_{i=N-k}^{N-1}} \lambda^2_i}}} & \textrm{if $\displaystyle{\sum_{i=N-k}^{N-1} \lambda^2_i\leq \sum_{i=N-k}^{N-1} \mu^2_i}$}\\
\\
{\sqrt{\frac{\displaystyle{\sum_{i=N-k}^{N-1}} (\lambda_i-\mu_i)^2}{\displaystyle{\sum_{i=N-k}^{N-1}} \mu^2_i}}}     & \textrm{if $\displaystyle{\sum_{i=N-k}^{N-1} \mu^2_i< \sum_{i=N-k}^{N-1} \lambda^2_i}$}\\
\end{cases}
$ & (\ref{eq:pincombe07detecting}) & \cite{pincombe07detecting}\\
\\
D2 & $\displaystyle{\epsilon(G,H) = \sqrt{\int_0^\infty \left[\rho_G(\omega)-\rho_H(\omega)\right]^2} \textrm{d}\omega}$
& (\ref{eq:ipsen02evolutionary}) & \cite{ipsen02evolutionary} \\
\\
D3 & $\displaystyle{d_M(G,H) = \sqrt{\sum_{i=0}^{N-1} \left(\lambda^{(G,M)}_i -\lambda^{(H,M)}_i\right)^2}}$
& (\ref{eq:zhu05study}) & \cite{zhu05study} \\
\\
D4 & $\displaystyle{d(G,H) = \sum_{i=0}^{N-1}\left(\lambda^{(G,L)}_i -\lambda^{(H,L)}_i\right)^2}$
& (\ref{eq:comellas08spectral}) & \cite{comellas08spectral}\\
\\
D5e & $\displaystyle{d(G,H) = \int_\lambda (1-\lambda)^4 \left(f_{\lambda,G}(\lambda)-f_{\lambda,H}(\lambda)\right)^2\textrm{d}{\lambda}}$
& (\ref{eq:fay09weighted_exact}) & \cite{fay09weighted} \\
\\
D5a & $\displaystyle{d_n(G,H) = \displaystyle{\sum_{k\in K} (1-k)^n (f_G(\lambda=k)- f_H(\lambda=k))^2}}$
& (\ref{eq:fay09weighted_approx}) & \cite{fay09weighted} \\
\\
D6 & $d\displaystyle{(G,H) = \sqrt{\textrm{JS}(f_G,f_H)}}$
& (\ref{eq:banerjee09structural}) & \cite{banerjee09structural} \\
\\
\hline
\end{tabular}
\end{center}
\end{table}

The last spectral measure D6 in this review was presented in \cite{banerjee09structural} and it employ two different divergence measures, Kullback-Leibler and Jensen-Shannon.
The Kullback-Leibler divergence measure is defined on two probability distributions $p_1$, $p_2$ of a discrete random variable $X$ as
\begin{displaymath}
\textrm{KL}(p_1,p_2) = \sum_{x\in X} p_1(x) \log\frac{p_1(x)}{p_2(x)}\ .
\end{displaymath}
The Kullback-Leibler divergence measure is not a metric, because is not symmetric and it does not satisfy the triangle inequality.
To overcome this problem, the author consider the Jensen-Shannon measure, which in some sense is the symmetrization of $\textrm{KL}$:
\begin{displaymath}
\textrm{JS}(p_1,p_2) = \frac{1}{2} \textrm{KL}\left(p_1,\frac{p_1+p_2}{2}\right) + \frac{1}{2} \textrm{KL}\left(p_2,\frac{p_1+p_2}{2}\right)\ .
\end{displaymath}
With this definition, the square root of $\textrm{JS}$ is a metric.
Thus, if $f$ is the (normalized Laplacian) spectral probability distribution, a distance between two networks can be defined as
\begin{equation}
\label{eq:banerjee09structural}
d(G,H) = \sqrt{\textrm{JS}(f_G,f_H)}\ .
\end{equation}

Clearly, all the above distances D1-D6 suffer from the existence of isospectral graphs: they are relatively rare (especially in real networks) and qualitatively similar.
For this reason, it would be more correct to call them distances between classes of isospectral networks.
The six described distances are analytically summarized in Tab. \ref{tab:distances}.

We conclude mentioning that spectrum of the graph can be indireclty used for assessing similarity \cite{robles-kelly03edit}. 
The authors employ a seriation method based on graph spectrum to convert the graph into a string so to get a sounder basis for the graph edit distance computation, aiming at the optimization of a function of the leading eigenvectors of the adjacency matrix.

\section{Benchmarking experiments}
\label{sec:results}
In this section, we demonstrate the use of the distances in
Tab. \ref{tab:distances} in the comparison of network topologies in a
controlled situation.  To such aim, we constructed three synthetic
benchmark datasets, detailed hereafter.  All simulations have been
performed within the R statistical environment \cite{r2009}.
Throughout all simulations, we kept, for each distance, the parameter
values as in the reference paper wherever possible, e.g.,
$\gamma=0.08$ for the scale of the Lorentz distribution in D2; the
heat diffusion kernel in D3; the time $t=3.5$ for the kernel in
distance D3.  For D1 we choose to use the $\lfloor\frac{N}{2}\rfloor$
largest eigenvalues.

\subsection{Data Description}
\label{ssec:data}

The simulated topologies are generated within the R statistical
environment \cite{r2009} by means of the simulator provided by the
package \textit{netsim} \cite{dicamillo07,dicamillo09gene}, producing
networks that reproduce principal characteristics of transcriptional
regulatory networks. The simulator takes into account the scale-free
distribution of the connectivity and constructs networks whose
clustering coefficient is independent of the number of nodes in the
network. All random graphs are generated by keeping the default
values of \textit{netsim} for the structural parameters.

\begin{figure}[h]
\begin{tabular}{cccc}
\includegraphics[width=2.8cm]{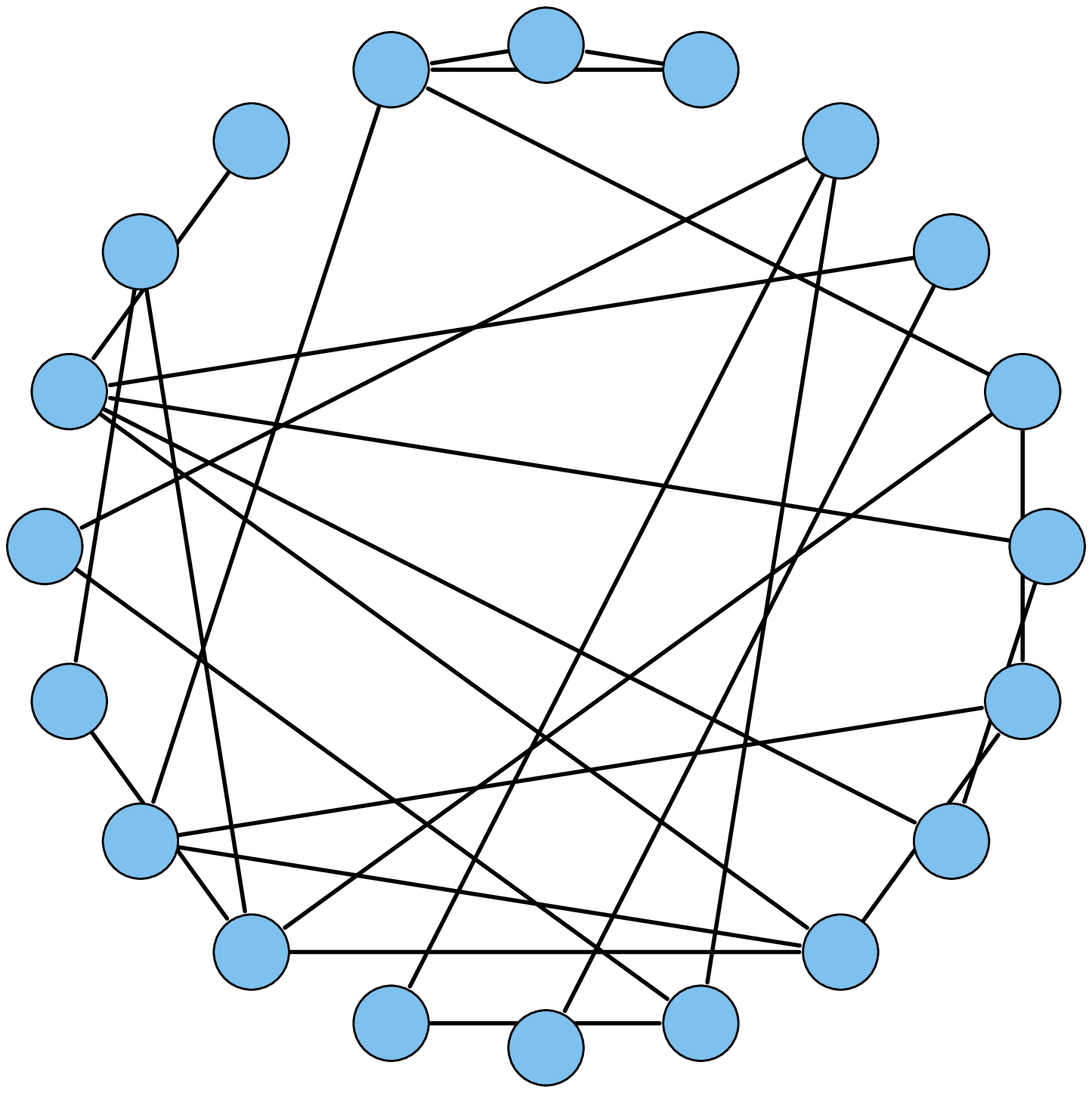}&
\includegraphics[width=2.8cm]{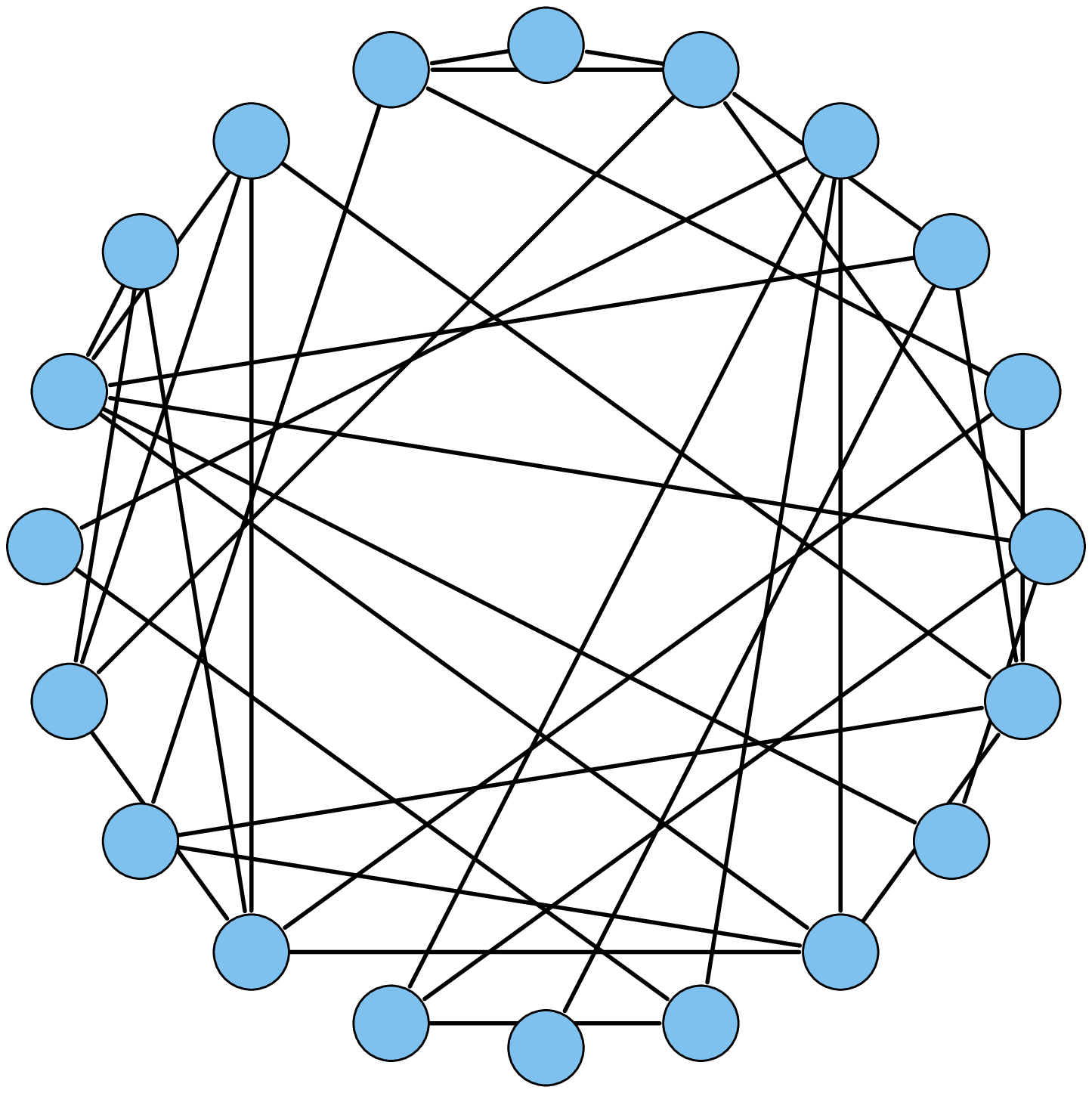}&
\includegraphics[width=2.8cm]{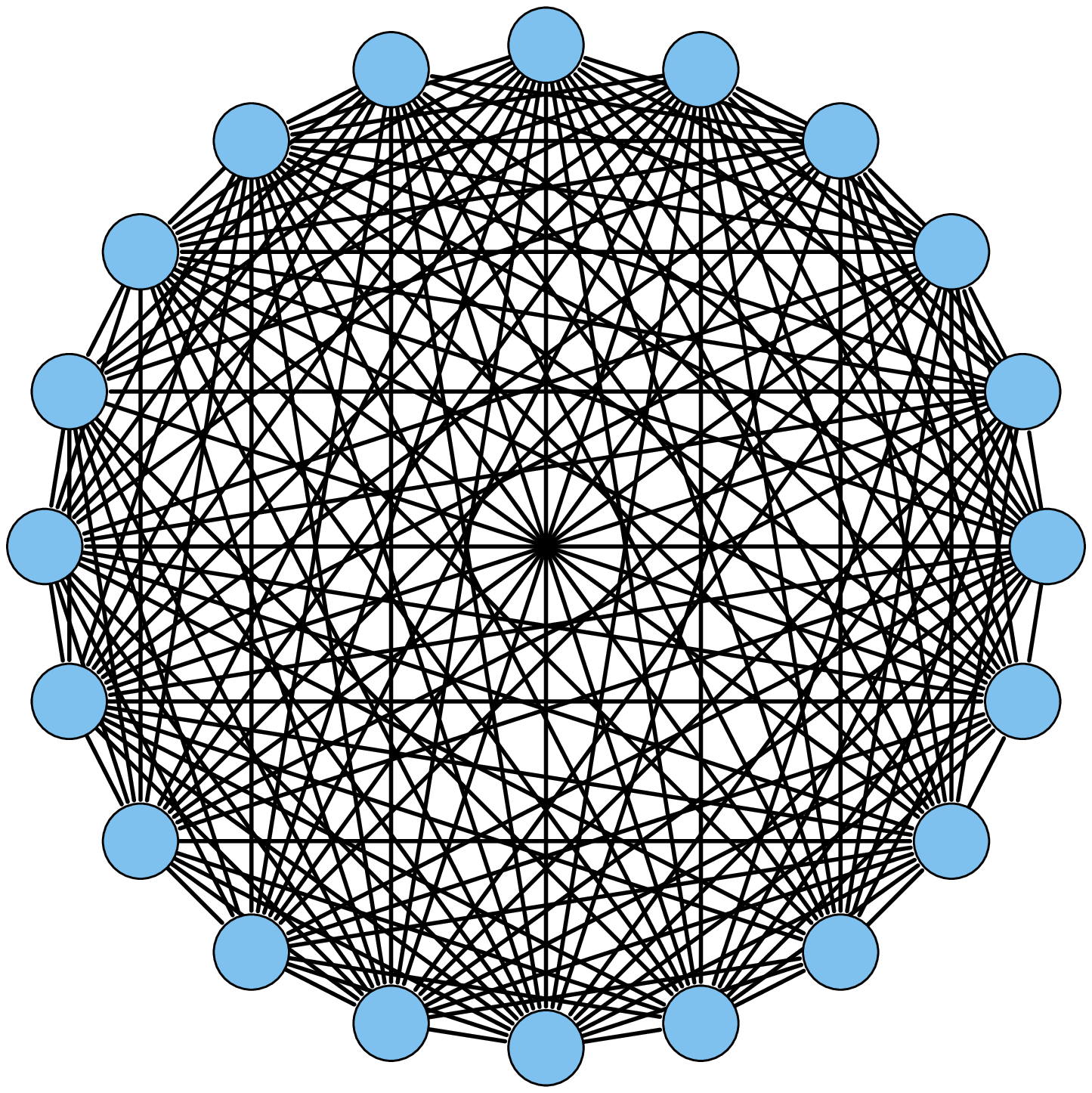}&
\includegraphics[width=2.8cm]{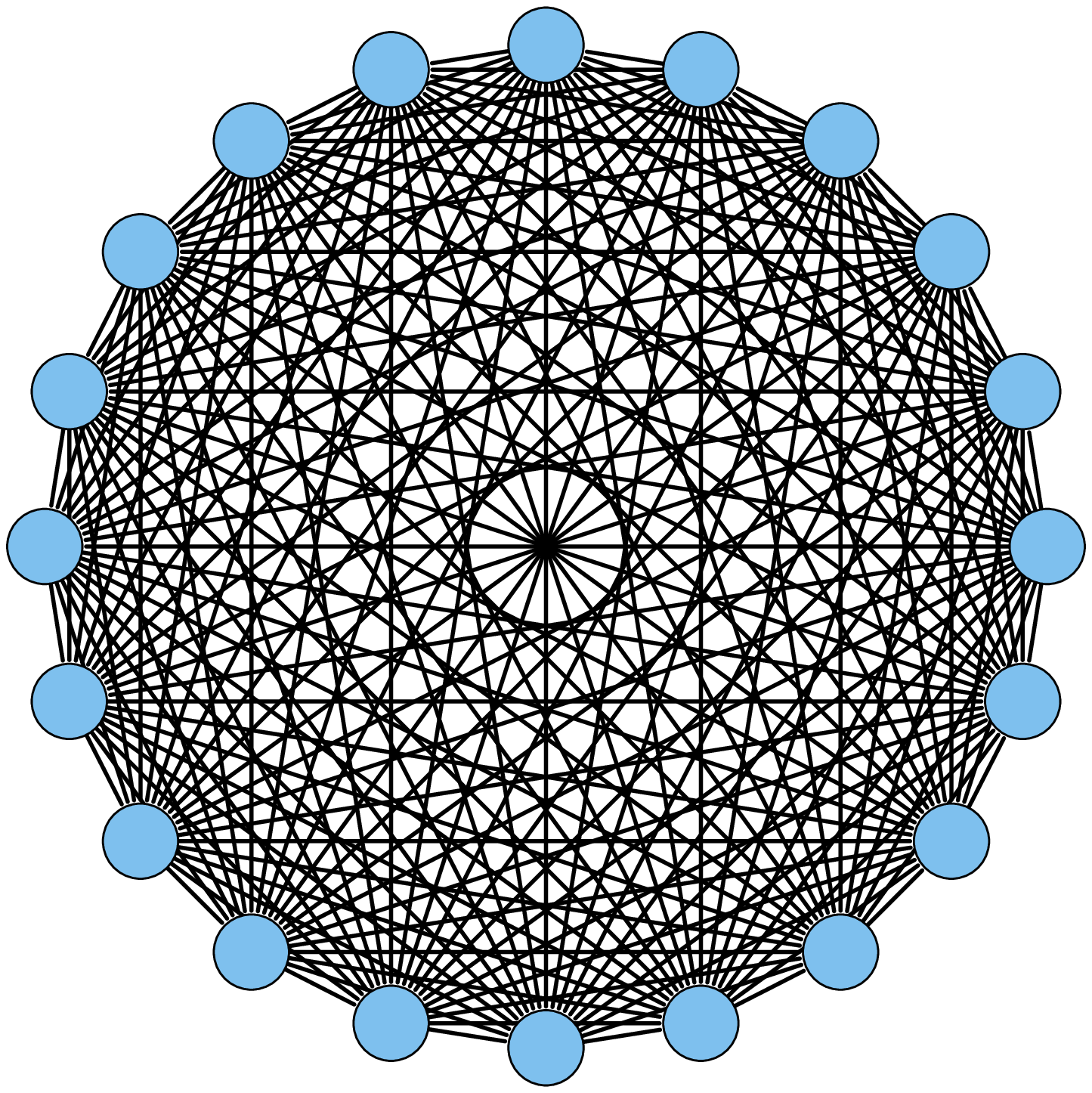}\\
$A$ & $A_5$ & $\overline{A}$ & $F$
\end{tabular}
\caption{Benchmark Dataset $\mathcal{B}_1(b, 25, 5)$: the original
graph $A$, the perturbed graph $A_5$, the complemental graph
$\overline{A}$ and the fully connected graph $F$.}
\label{fig:bench1}
\end{figure}

In the first experiment we consider a random network $A$ on $N$
vertices and we compare it with the full connected network with the
same number of nodes $F$, the complemental network $\overline{A}$ and
a matrix $A_{p}$ obtained from $A$ by modifying (inserting/deleting)
about the $p$\% of the nodes.  For smoothing purposes, the process is
repeated $b$ times to obtain the first benchmarking dataset
$\mathcal{B}_1(b,N,p)$.  An instance of this benchmark dataset is
shown in Fig.\ref{fig:bench1}.
In Tab. ~\ref{tab:nodes_1} we show the average on $b=50$ instances of
the number of nodes of the starting matrix $A$ and the perturbed
matrix $A_5$.  Because of the small number of links in the original
matrix, the 5\% perturbation mostly reflects in links insertion. On
average, the density of the original graph $A$ can be expressed by the
relation $l\simeq 1.7N-5$, where $l$ is the number of links and $N$
the number of vertices.

\begin{table}[b]
\caption{Number of links in the original matrix $A$, in the fully
connected matrix $F$ (maximum number of links for the given dimension)
and in the perturbed matrix $A_5$, expressed as mean $\pm$ standard
deviation on 50 replicates.}
\label{tab:nodes_1}
\begin{center}
\begin{tabular}{cccc}
\hline
$\mathbf{N}$ & $\mathbf{F}$ & $\mathbf{A}$ & $\mathbf{A_5}$ \\
\hline
10 & 45 & 13.4$\pm$2.0 & 13.1$\pm$2.3\\
20 & 190 & 29.0$\pm$3.6 & 36.6$\pm$5.2 \\
50 & 1225 & 79.3$\pm$7.4 & 131.8$\pm$4.2\\
100 & 4950 & 164.5$\pm$13.6 & 388.2$\pm$12.1\\
\hline
\end{tabular}
\end{center}
\end{table}

\begin{figure}[t]
\begin{tabular}{ccc}
\includegraphics[width=4cm]{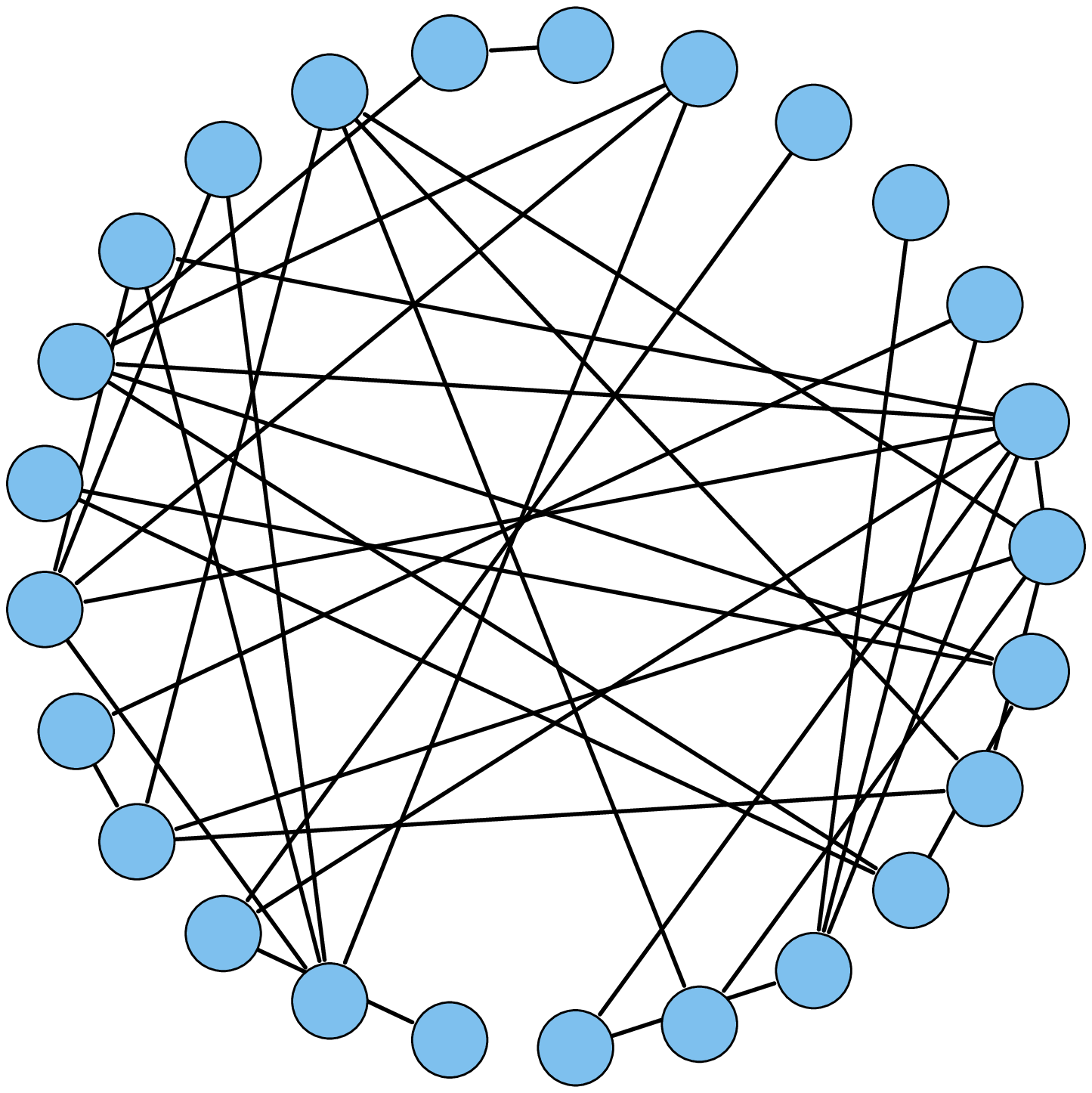}&
\includegraphics[width=4cm]{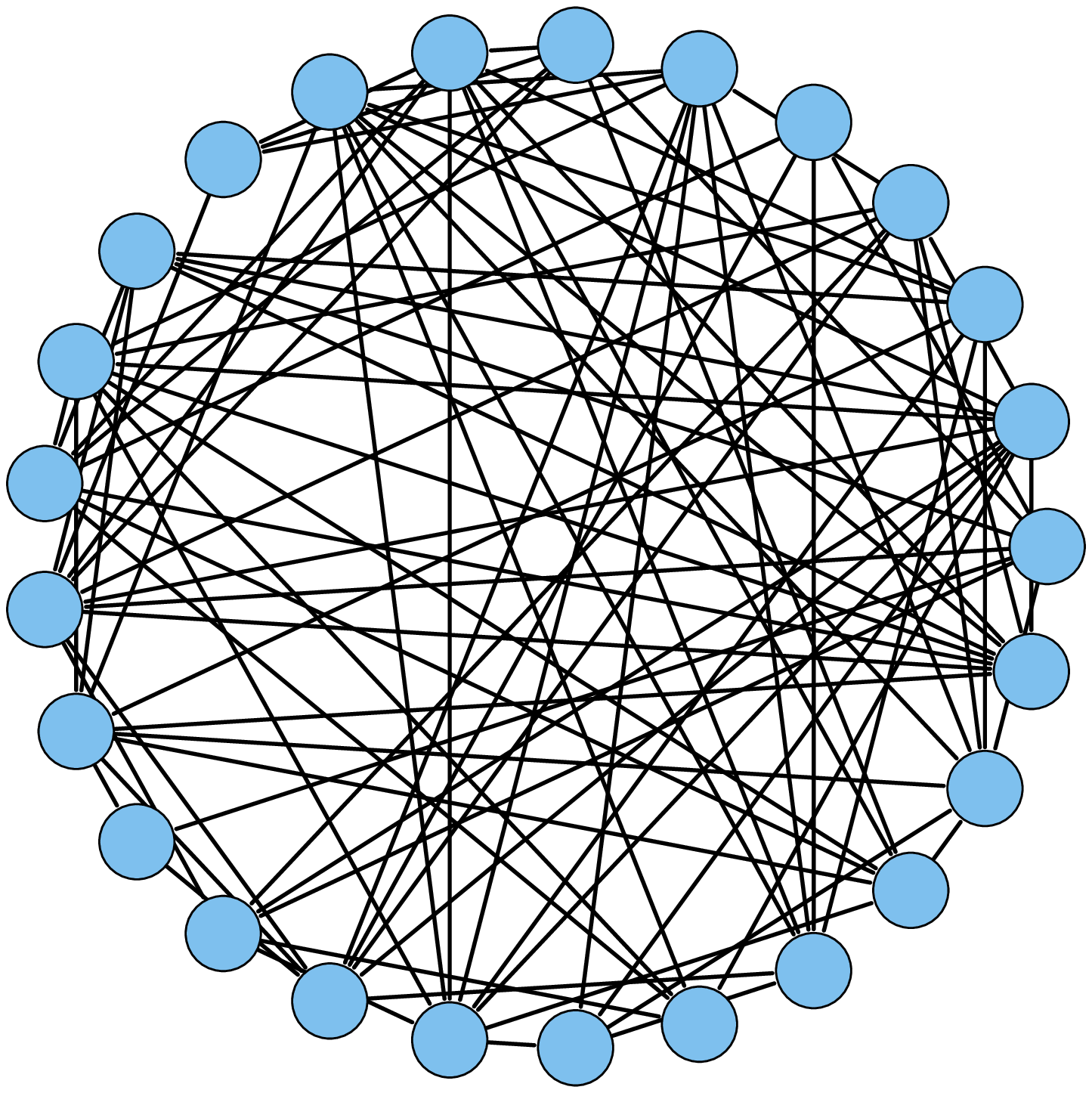}&
\includegraphics[width=4cm]{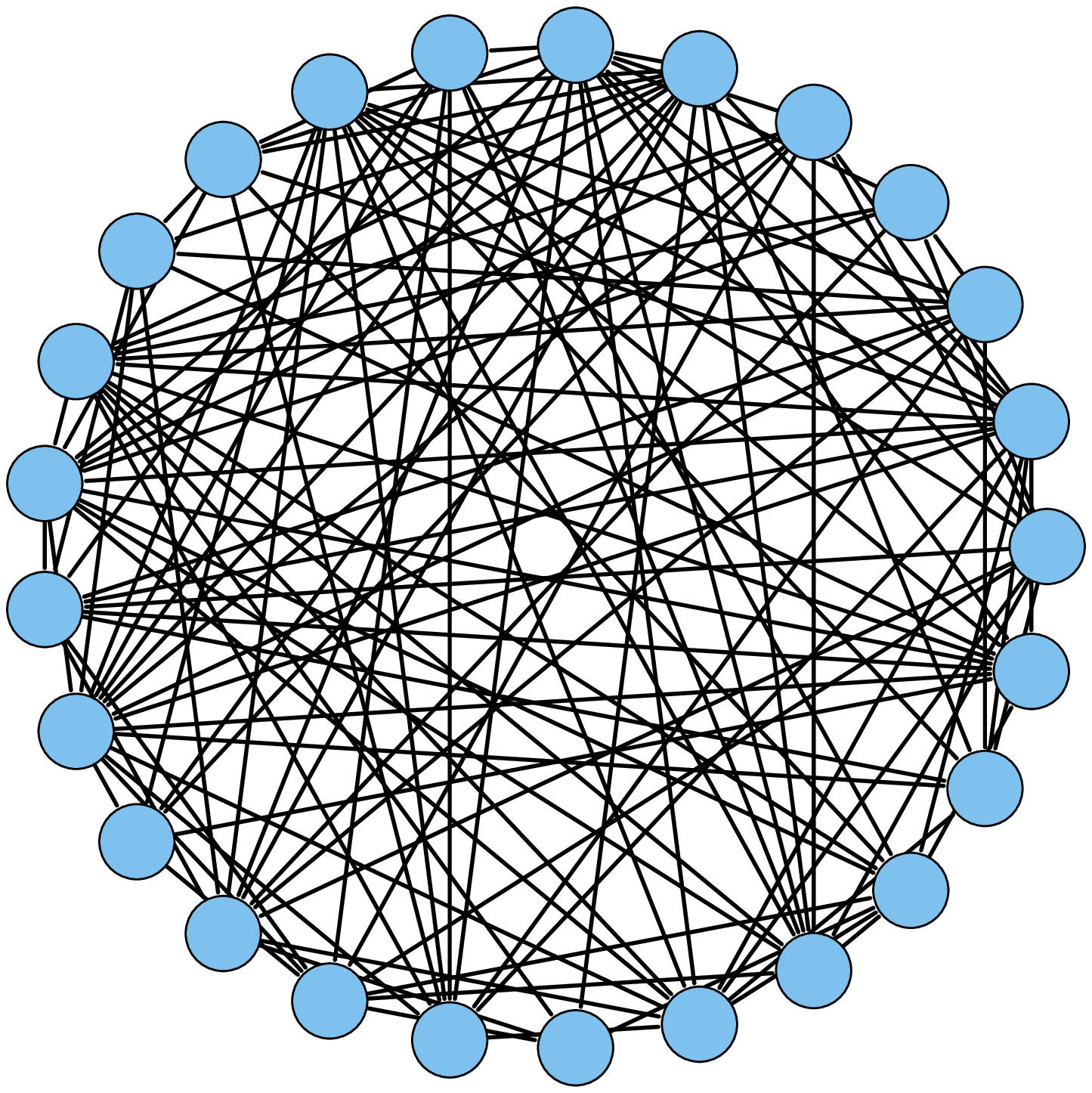}\\
$S_1$ & $S_{10}$ & $S_{20}$ 
\end{tabular}
\caption{Benchmark Dataset $\mathcal{B}_2(b,20,25,5)$: the original graph $S_1$ (first element of the series), the tenth element $S_{10}$ of the series and the final graph $S_{20}$.}
\label{fig:bench2}
\end{figure}

In the second experiment we simulate a time-series of $T$ networks on
$N$ nodes starting from a randomly generated graph $S_1$, where each
successive element $S_i$ of the series is generated from its ancestor
$S_{i-1}$ by randomly modifying $p$\% of the links.  Again $b=50$
instances of the series are created and collected into the second
benchmarking dataset $\mathcal{B}_2(b,T,N,p)$.  With this strategy,
the number of existing links is increasing with the series index,
being the original adjacency matrix almost sparse.  The starting
matrix $S_1$ has on average 38.1$\pm$5.2 nodes, while the last element
of the series $S_{20}$ has 132.3$\pm$8.2.  Three elements of this
benchmark dataset are shown in Fig.\ref{fig:bench2}.

\begin{figure}[b]
\begin{tabular}{ccc}
\includegraphics[width=4cm]{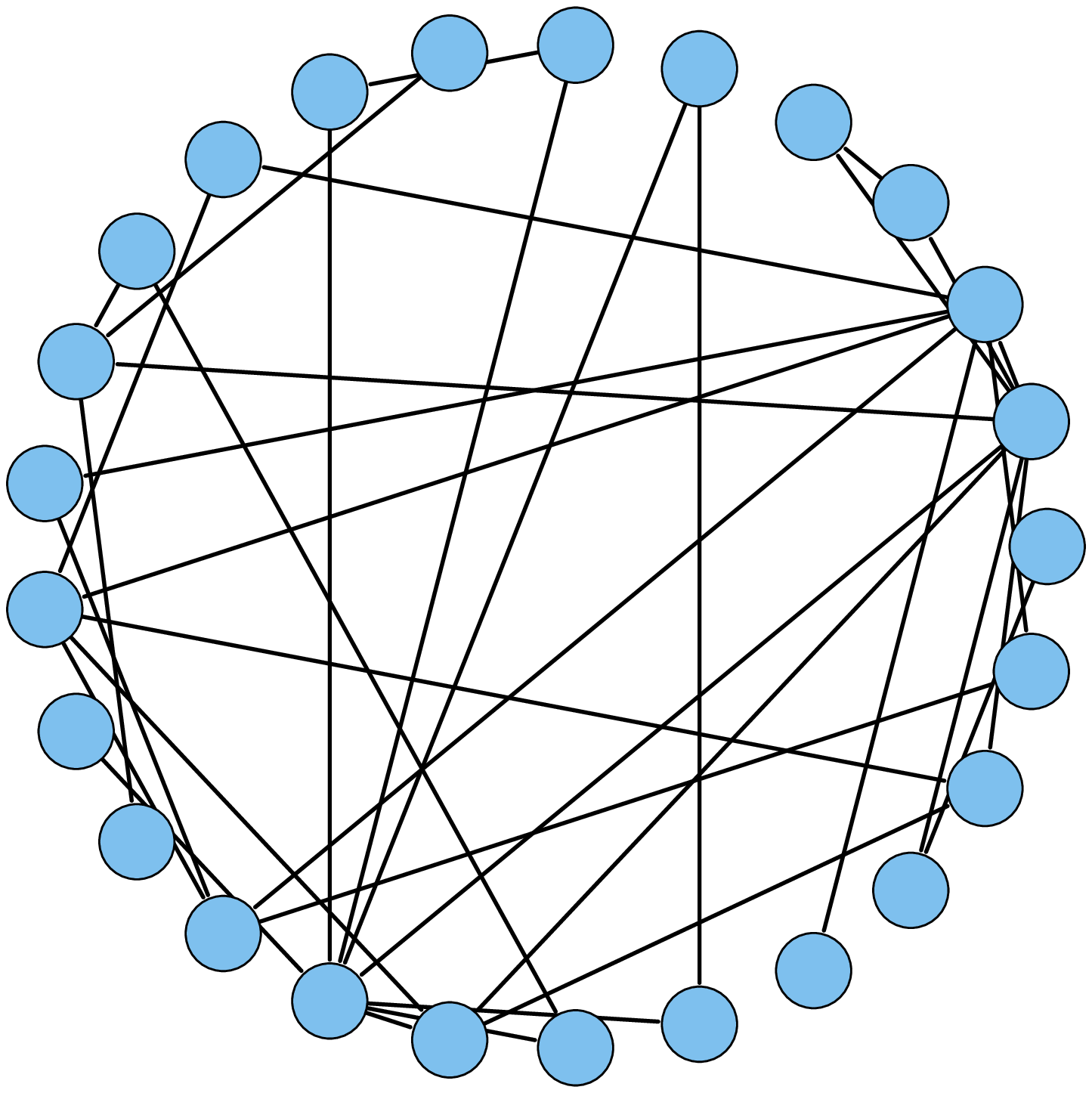}&
\includegraphics[width=4cm]{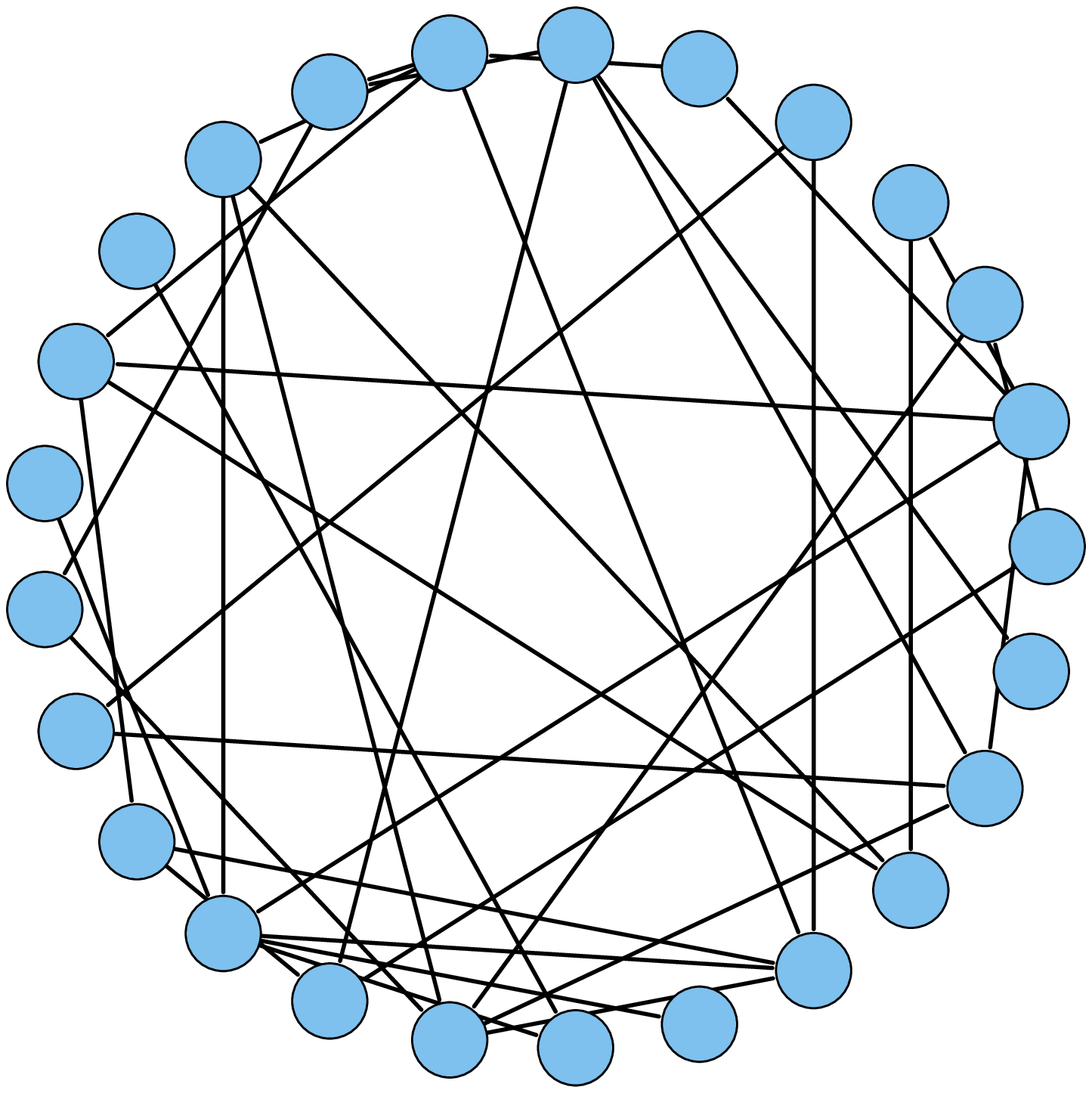}&
\includegraphics[width=4cm]{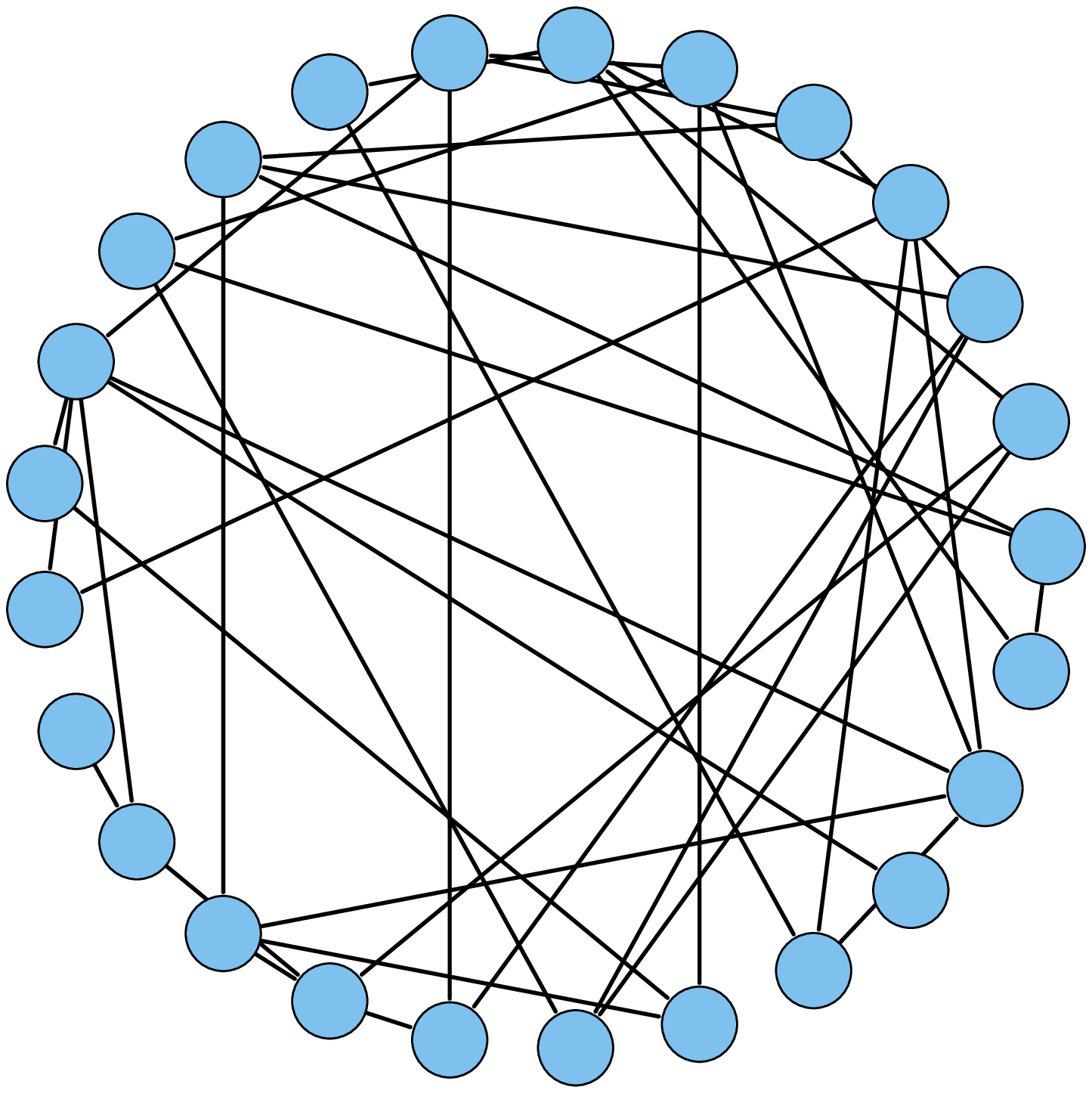}\\
$S_1$ & $S_{10}$ & $S_{20}$
\end{tabular}
\caption{Benchmark Dataset $\mathcal{B}_3(b,20,25,5,5)$: the original graph $S_1$ (first element of the series), the tenth element $S_{10}$ of the series and the final graph $S_{20}$.}
\label{fig:bench3}
\end{figure}

The third experiment is based on a benchmark dataset $\mathcal{B}_3(b,T,N,nd,na)$.
Starting from $\mathcal{B}_2(b,T,N,p)$, different perturbations are applied: each successive element $S_i$ of the series is generated from its ancestor $S_{i-1}$ by randomly deleting $nd$ links and adding $na$ links. 
By construction, the number of existing links for all elements of the series is constant. 
Three elements of $\mathcal{B}_3(b,20,25,5,5)$ are shown in Fig.\ref{fig:bench3}.

\subsection{Results}
\label{ssec:results}

In Exp. 1 the six distances D1-D6 were applied on 4 instances of $\mathcal{B}_1(50,N,5)$ for $N=10,20,25,100$ and distances between the original graph $A$ and the three companion matrices $F$, $\overline{A}$ and $A_p$ were computed.
Results are collected in Tab. \ref{tab:exp1}.

\begin{table}
\caption{Results of the experiments on the first benchmarking dataset. For each measure D1-D6 and number of network vertices $N$, the values are reported of the distances between the network $A$ and the networks $A_{5}$, $\overline{A}$ and $F$ in terms of the minimum (m), mean ($\mu$) $\pm$ standard deviation and maximum (M) on the 50 replicates. Values of D5 are in $10^{-3}$.}
\label{tab:exp1}
\begin{center}
\scriptsize
{\tiny
\begin{tabular}{r|r|rrr|rrr|rrr}
\hline
$\mathbf{N}$ & $\mathbf{D}$ & \multicolumn{3}{c|}{$\mathbf{A_5}$} & \multicolumn{3}{c|}{$\mathbf{\overline{A}}$} & \multicolumn{3}{c}{$\mathbf{F}$} \\
 & & m & $\mu \pm \sigma$ & M & m & $\mu \pm \sigma$ & M & m & $\mu \pm \sigma$ & M \\
\hline
10 & 1 & 0.025 & 0.108 $\pm$ 0.053 & 0.197 &  0.085 & 0.982    $\pm$ 0.383 & 1.564 &  0.424 & 1.324 $\pm$  0.350 & 1.811  \\ 
10 & 2 & 0.215 & 0.319 $\pm$  0.052 & 0.403 &  0.47 & 0.857    $\pm$ 0.174 & 1.066 &  1.434 & 1.563  $\pm$  0.04 & 1.635  \\ 
10 & 3 & 0 & 0.067 $\pm$ 0.074 & 0.294 &  0.006 & 0.415  $\pm$ 0.39 & 1.83 & 0.028 & 0.472  $\pm$  0.402 & 1.925 \\ 
10 & 4 & 0& 2.182  $\pm$ 1.01  &4.533 &14.33 &151.8  $\pm$ 71.5 &328.1 &336 &470.4  $\pm$ 61.7 & 598\\ 
10 & 5 & 0 & 0.941   $\pm$ 0.603 & 1.844 &  0.092 & 3.635  $\pm$  2.340 & 8.907 &  0.518 & 4.112   $\pm$  2.306 & 9.491  \\ 
10 & 6 & 0.102 & 0.169 $\pm$ 0.039 & 0.259 &  0.192 & 0.386    $\pm$  0.084 & 0.507 &  0.431 & 0.507  $\pm$  0.04 & 0.552  \\ 
\hline
20 & 1 & 0.037 & 0.194 $\pm$ 0.069 & 0.342 &  2.117 & 2.768  $\pm$ 0.379 &3.71 &  2.455 & 3.038 $\pm$  0.372 & 4.006  \\ 
20 & 2 & 0.202 & 0.284 $\pm$ 0.049 & 0.381 &  1.025 & 1.091  $\pm$  0.034 &1.165 &  1.538 & 1.55 $\pm$   0.008 & 1.568  \\ 
20 & 3 & 0.044 & 0.154 $\pm$ 0.132 & 0.577 &  0.588 & 1.04  $\pm$  0.333 &2.05 &  0.643 & 1.103 $\pm$  0.336 & 2.123  \\ 
20 & 4 & 1.812& 15.9$\pm$ 6.5& 28.5& 2584&3658  $\pm$ 420 &4761 &4898 &5531  $\pm$ 243&6146 \\ 
20 & 5 & 0.358 & 0.836 $\pm$ 0.503 & 2.459 & 2.416 & 3.623  $\pm$  6.441 &      1.041 & 2.439 & 3.654  $\pm$  6.45   & 1.036 \\ 
20 & 6 & 0.135 & 0.207 $\pm$ 0.04 & 0.323 &  0.581 & 0.772  $\pm$  0.879 & 0.077 & 0.652 & 0.767  $\pm$  0.83 & 0.05 \\ 
\hline
50 & 1 & 0.389 & 0.504 $\pm$ 0.072 & 0.606 &  6.676 & 8.057  $\pm$ 0.784 & 9.064 &  6.924 & 8.288  $\pm$  0.771 & 9.253  \\ 
50 & 2 & 0.275 & 0.344 $\pm$ 0.042 & 0.437 &  1.152 & 1.195  $\pm$ 0.025 & 1.228 &  1.533 & 1.54 $\pm$    0.005 & 1.549  \\ 
50 & 3 & 0.668 & 1.186  $\pm$ 0.313 & 1.77 &  2.078 & 3.356  $\pm$    0.647 & 4.428 &  2.138 & 3.423 $\pm$   0.649 & 4.497  \\ 
50 & 4 & 138& 237 $\pm$ 48 &353 &83850 &92670  $\pm$ 3078 & 97710 &102700  &107300  $\pm$ 1613&110000\\ 
50 & 5 & 0.888 & 1.875 $\pm$ 0.541 &2.765 &  2.379 & 3.993    $\pm$ 0.847 & 5.42 &  2.379 & 3.992  $\pm$   0.849 & 5.42  \\ 
50 & 6 & 0.435 & 0.559 $\pm$ 0.0751 & 0.711 &  1.372 & 1.481  $\pm$ 0.061 & 1.597 &  1.183 & 1.277  $\pm$  0.063 & 1.39  \\ 
\hline
100 & 1 & 0.804 & 0.977 $\pm$ 0.076 & 1.086  &  13.55 & 16.07   $\pm$ 1.032   & 17.6   &  13.77 & 16.28   $\pm$  1.027 & 17.8  \\ 
100 & 2 & 0.451 & 0.506 $\pm$ 0.025 & 0.544  &  1.215 & 1.264  $\pm$ 0.019    & 1.293  &  1.524 & 1.533  $\pm$  0.004 & 1.543  \\ 
100 & 3 & 2.116 & 3.606 $\pm$ 0.665 & 4.64   &  4.506 & 6.723     $\pm$ 0.992 & 8.166  &  4.566 & 6.79   $\pm$  0.995 & 8.238  \\ 
100 & 4 & 1784  & 2161  $\pm$ 136    &240    & 842900 &861200  $\pm$ 9575      & 880600 &915800  &925100  $\pm$ 4880   & 935900\\ 
100 & 5 & 1.645 & 2.787 $\pm$ 0.525  & 3.589 &  2.602 & 3.941  $\pm$ 0.630     & 4.824  &  2.602 & 3.941 $\pm$  0.631 & 4.824 \\ 
100 & 6 & 0.933 & 1.102 $\pm$ 0.074  & 1.204 &  2.07 & 2.229     $\pm$ 0.088   & 2.397  &  1.694 & 1.839 $\pm$  0.078 & 1.997  \\ 
\hline
\end{tabular}
}
\normalsize
\end{center}
\end{table}

Distance D4 spans a considerably wider range than other measures, due
to the absence of the square root in the comparison of the Laplacian
spectra, while D5 is restricted into a very small interval. The same
distance D4 also shows a high dependency on the dimension of the
considered matrices and the number of the links (see Tab. \ref{tab:exp1}).

The best stability in terms of the relative standard deviation
$\sigma / \mu$ is reached by D2 and D4. Furthermore, D2,
differently from all other measures, is almost independent of the
number of vertices. Finally, D6 is the only measure that, in the cases
with $N > 10$, gives a lower distance for $F$ than for
$\overline{A}$.

\begin{figure}[!t]
\begin{center}
{\tiny
\begin{tabular}{ccc}
\includegraphics[width=3.2cm]{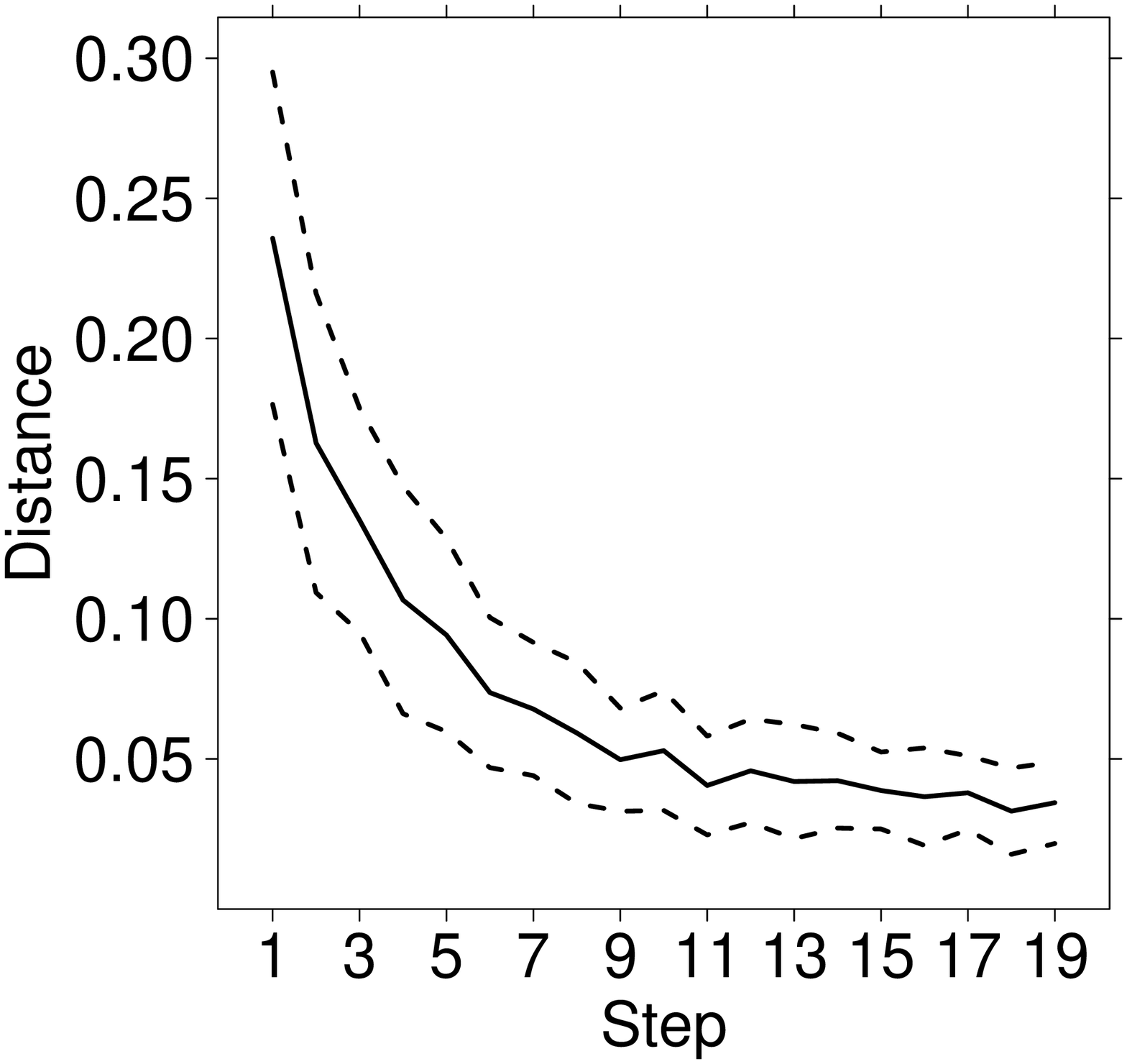}&
\includegraphics[width=3.2cm]{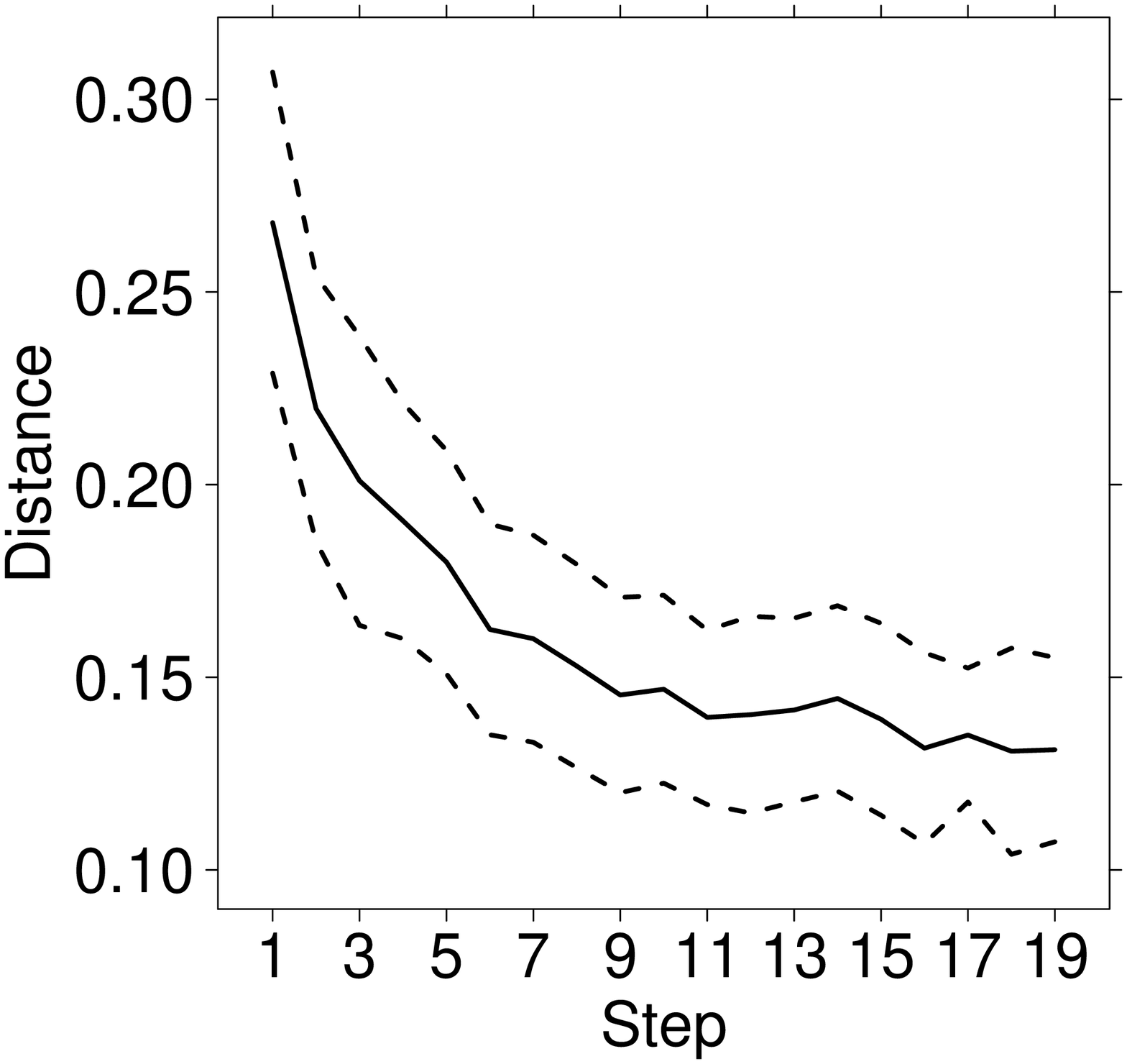}&
\includegraphics[width=3.2cm]{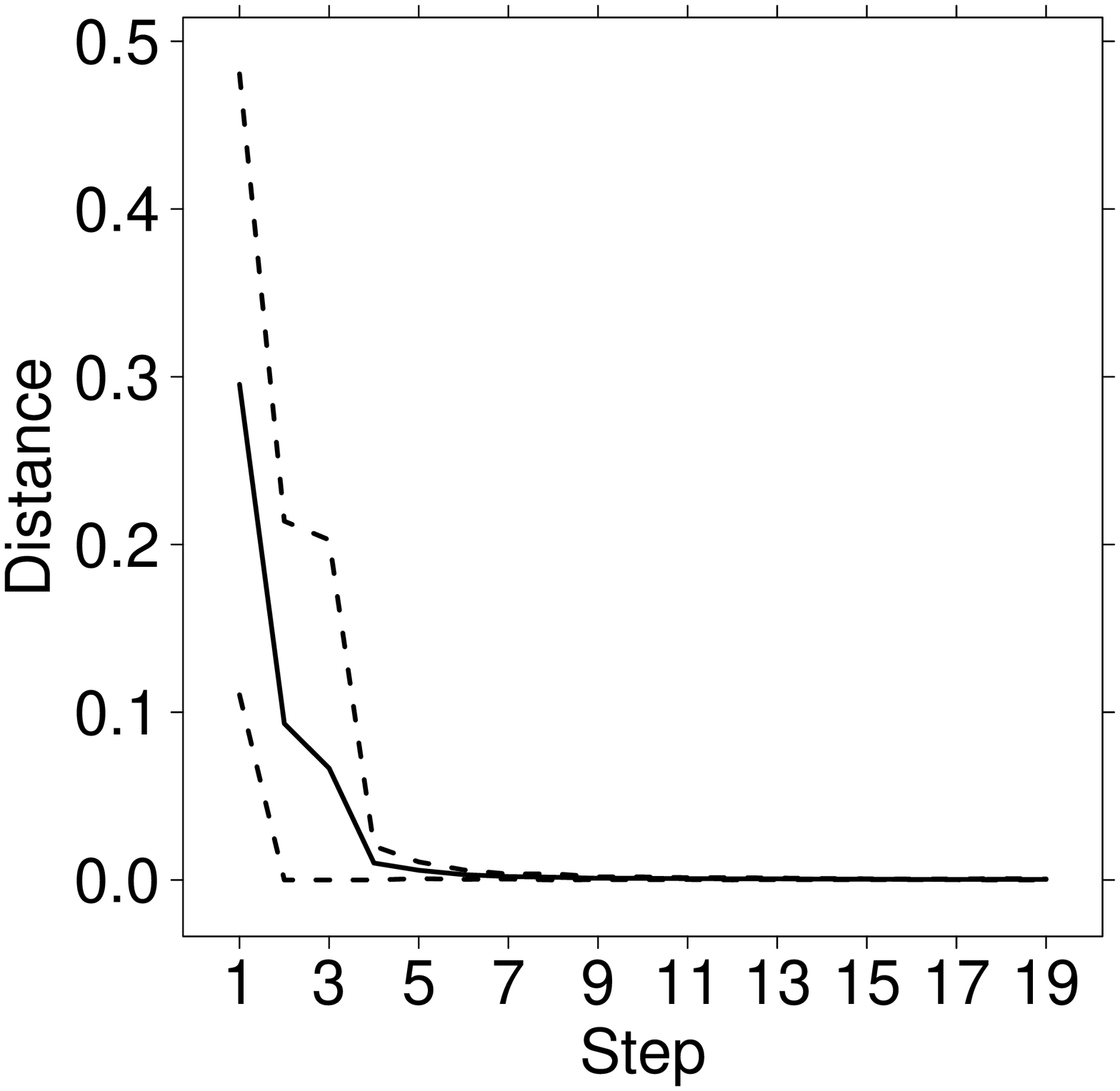}\\
D1 & D2 & D3\\
\includegraphics[width=3.2cm]{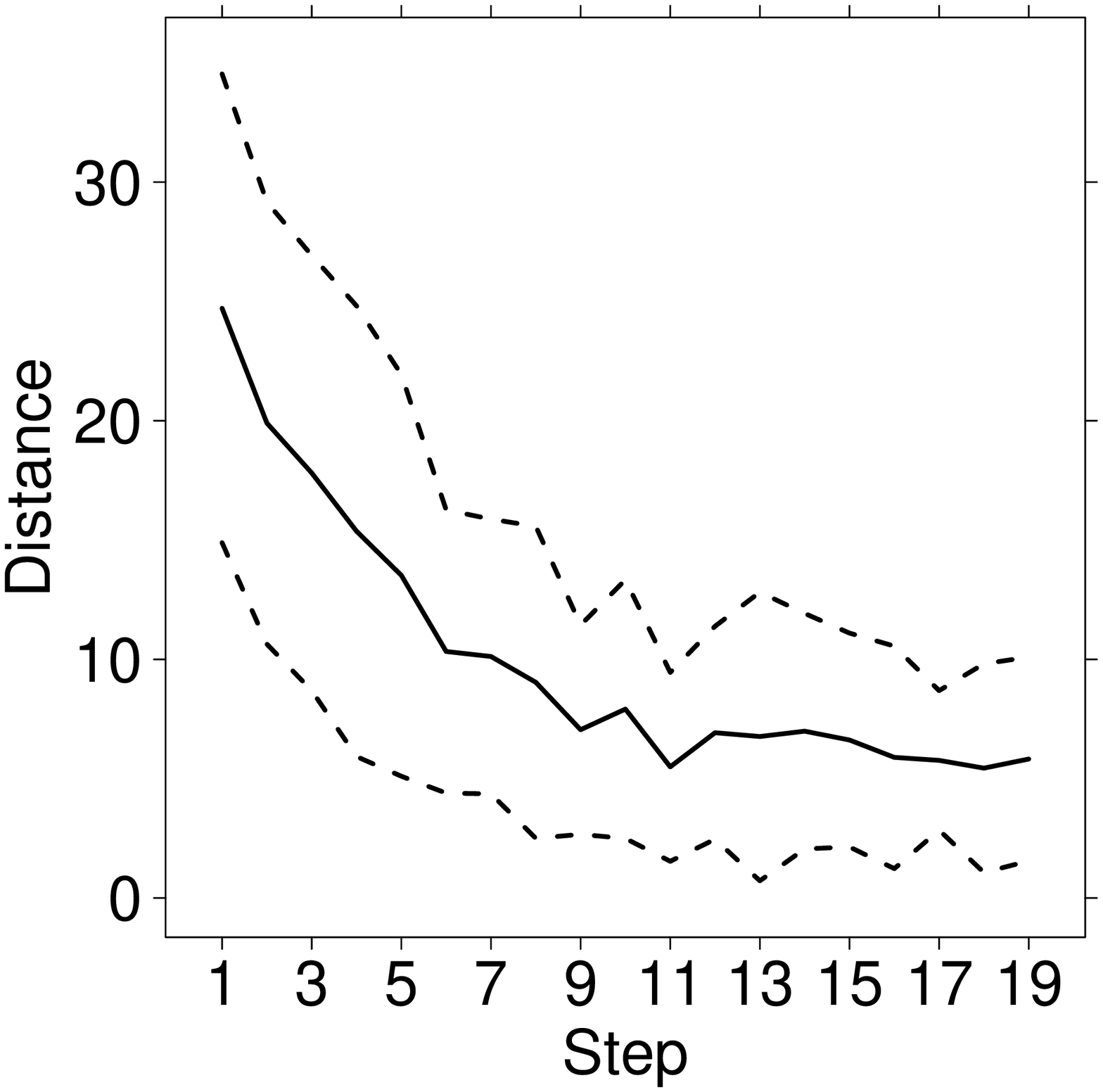}&
\raisebox{-0.2cm}{\includegraphics[width=3.6cm]{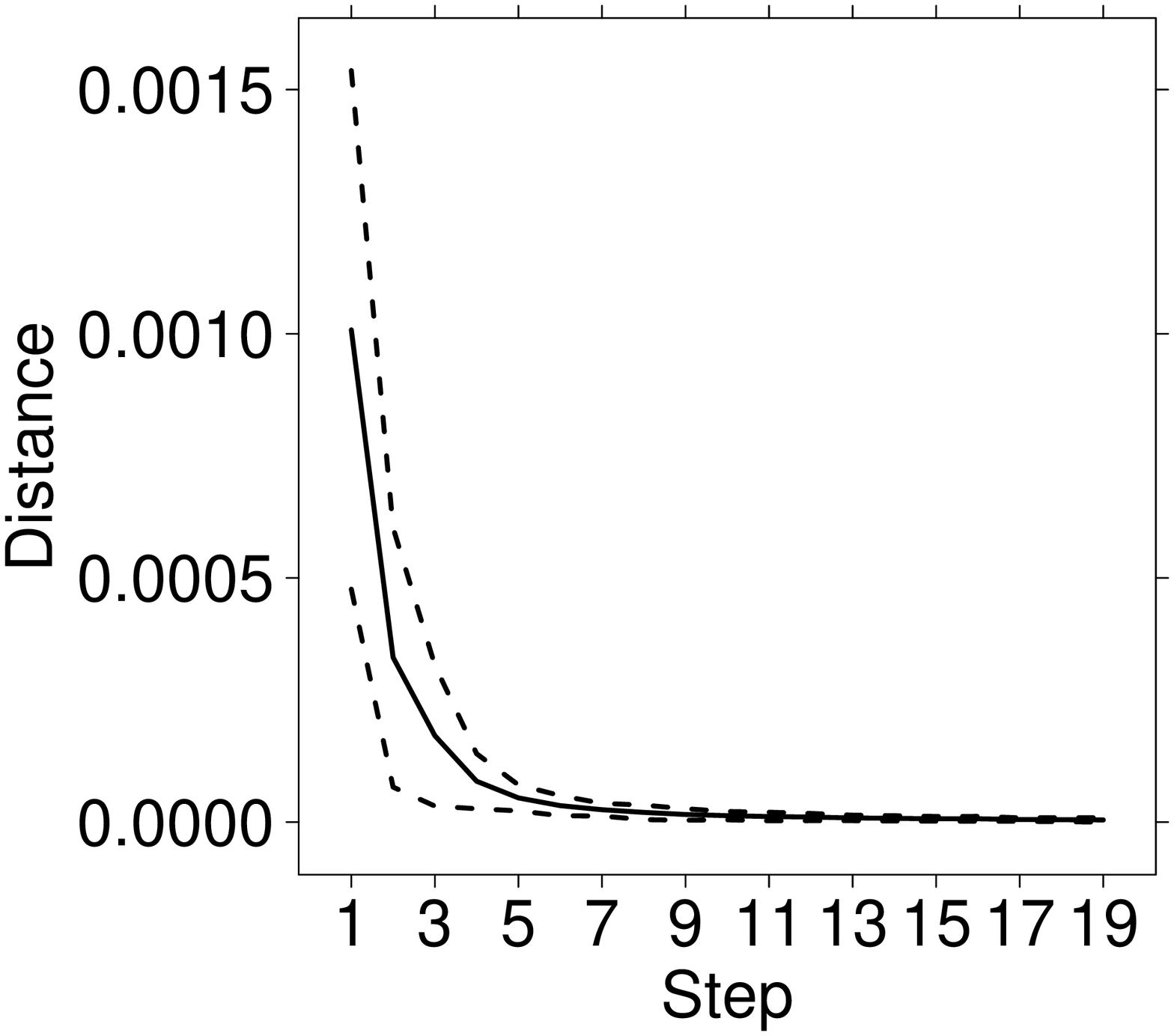}}&
\includegraphics[width=3.3cm]{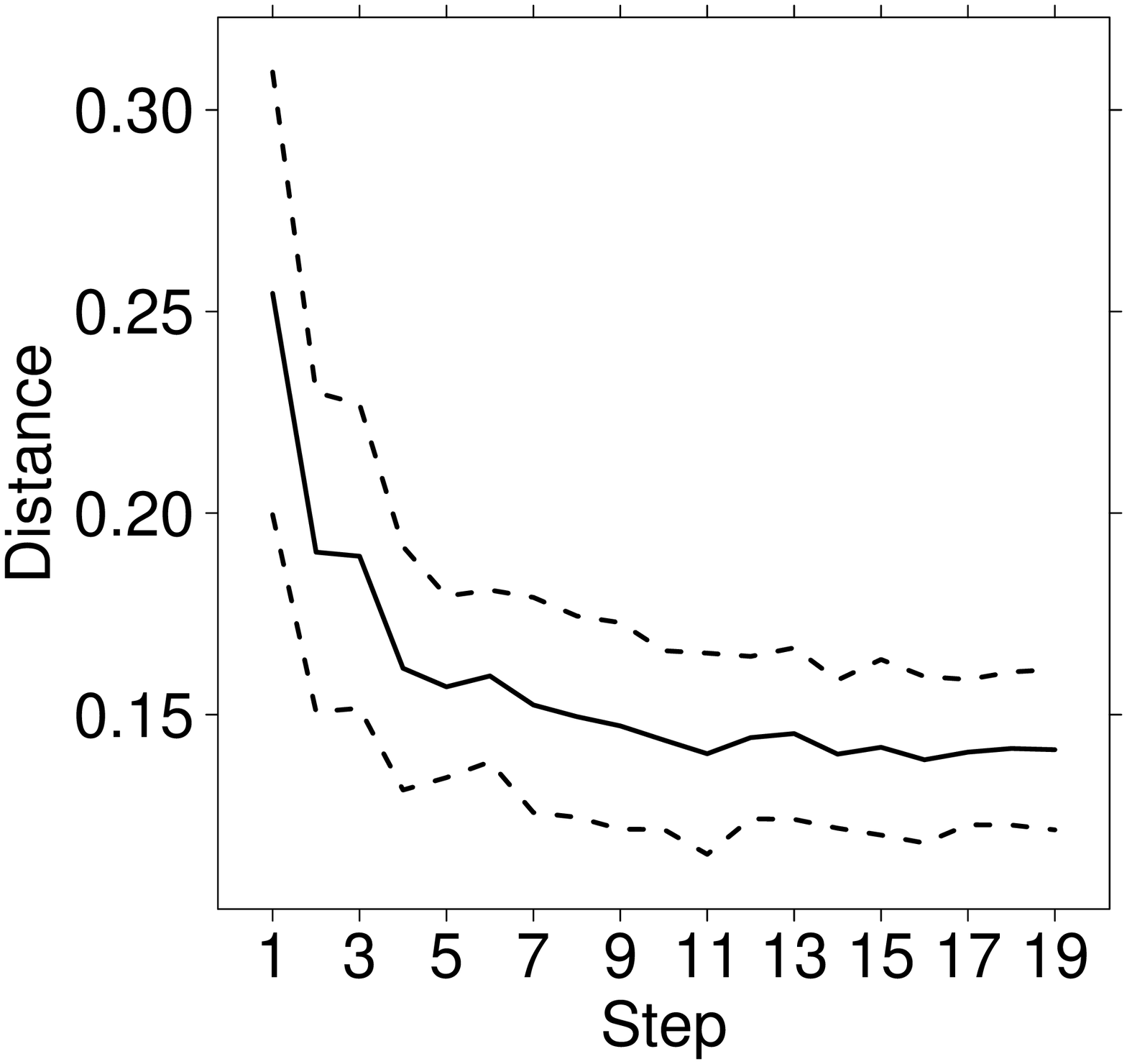}\\
D4 & D5 & D6 \\
\end{tabular}
}
\end{center}
\caption{Plots of the distances of consecutive elements of the series
of the dataset $\mathcal{B}_2(50,20,25,5)$. Solid line: mean
over the $b=50$ replicates; dashed lines: $1\sigma$ standard
deviation confidence intervals.}
\label{fig:exp2_trends}
\end{figure}

The summary plots in Fig. \ref{fig:exp2_trends} display results of
Exp. 2 on the benchmark dataset $\mathcal{B}_2(50,20,25,5)$.
Distances between consecutive elements $(S_i,S_{i+1})$ of the series
(defined Step $i$) were computed: results are averaged on the $50$
replicates.  
For all D1-D6, distance decreases for increasing steps, although on different ranges
(as already pointed out for Experiment 1) and with different widths for
the confidence intervals. 
D3 and D5 decrease more quickly for initial steps, so they are less
useful when comparing large networks.

\begin{figure}[!ht]
\begin{center}
\includegraphics[width=0.75\textwidth]{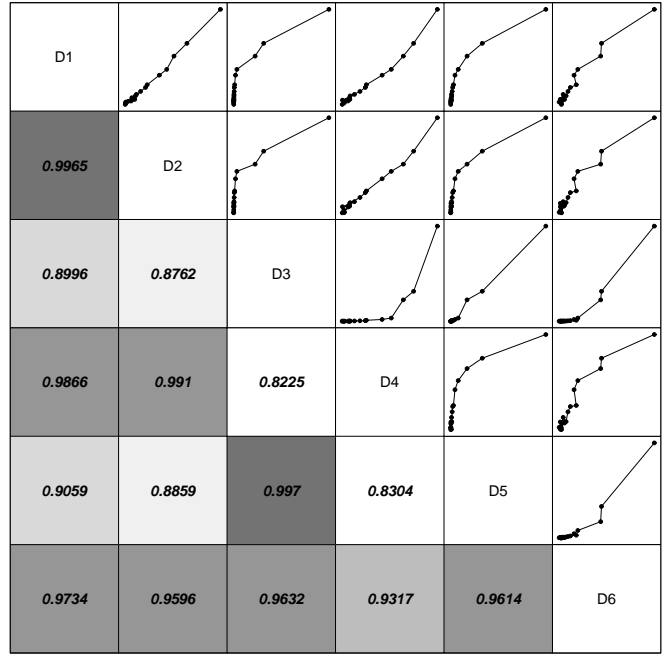}
\end{center}
\caption{Mutual scatterplots (upper triangle) and correlation values (lower triangle) for the Exp. 2.}
\label{fig:exp2_splom}
\end{figure}

To better highlight similarities and differences among the distances
regardless of their ranges of values, we also computed their mutual
correlations and plotted the mutual scatter plots in
Fig. \ref{fig:exp2_splom}.  All correlation values are quite high,
ranging from 0.8225 to 0.9970: D3 and D5 are mutually strongly
correlated, but they tend to separate from the other distances, as
evidenced both from the global correlation values and the scatter plot
profiles distancing from the panel diagonals.

\begin{figure}[!t]
\begin{center}
{\tiny
\begin{tabular}{ccc}
\includegraphics[width=3.2cm]{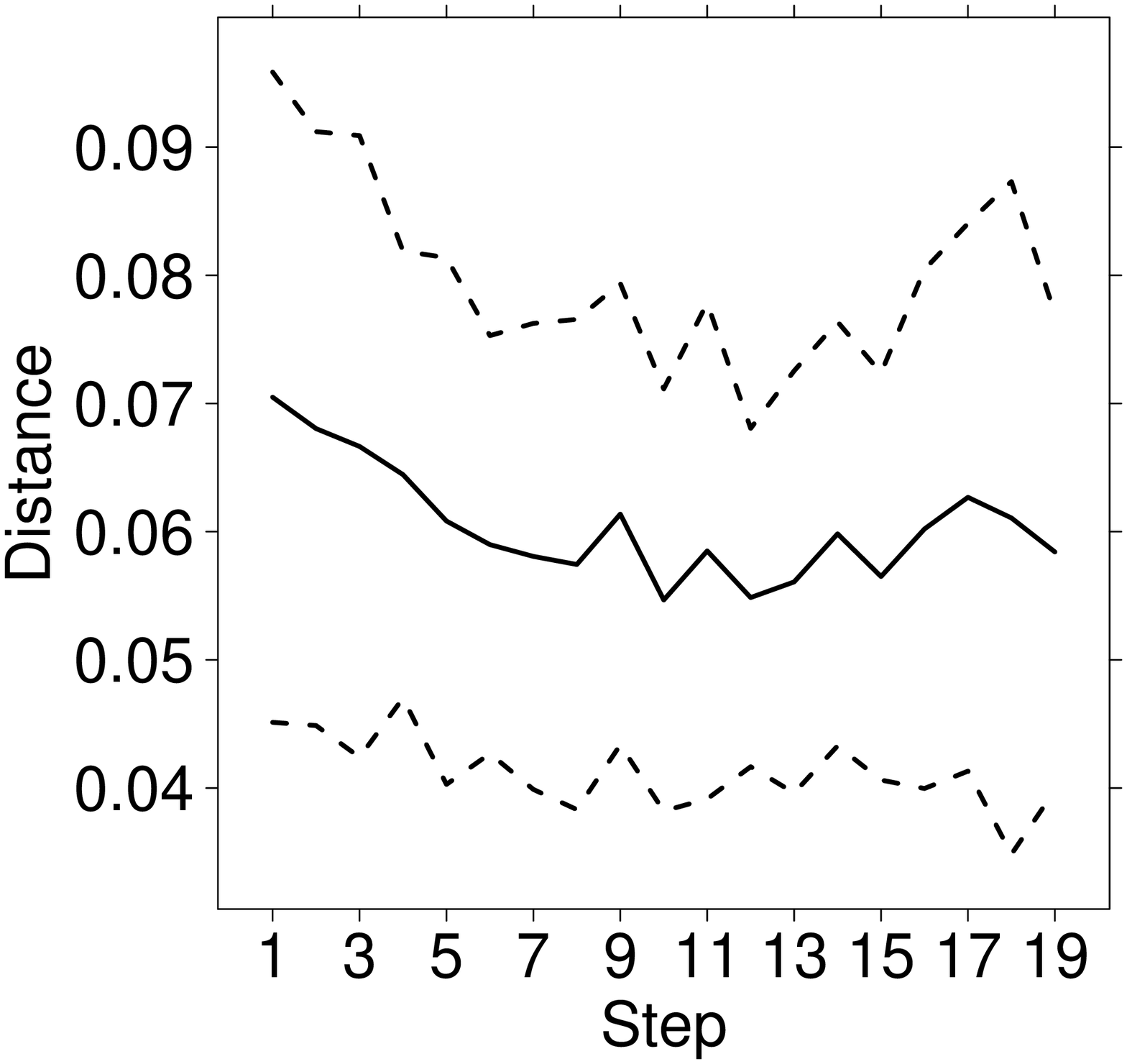}&
\includegraphics[width=3.2cm]{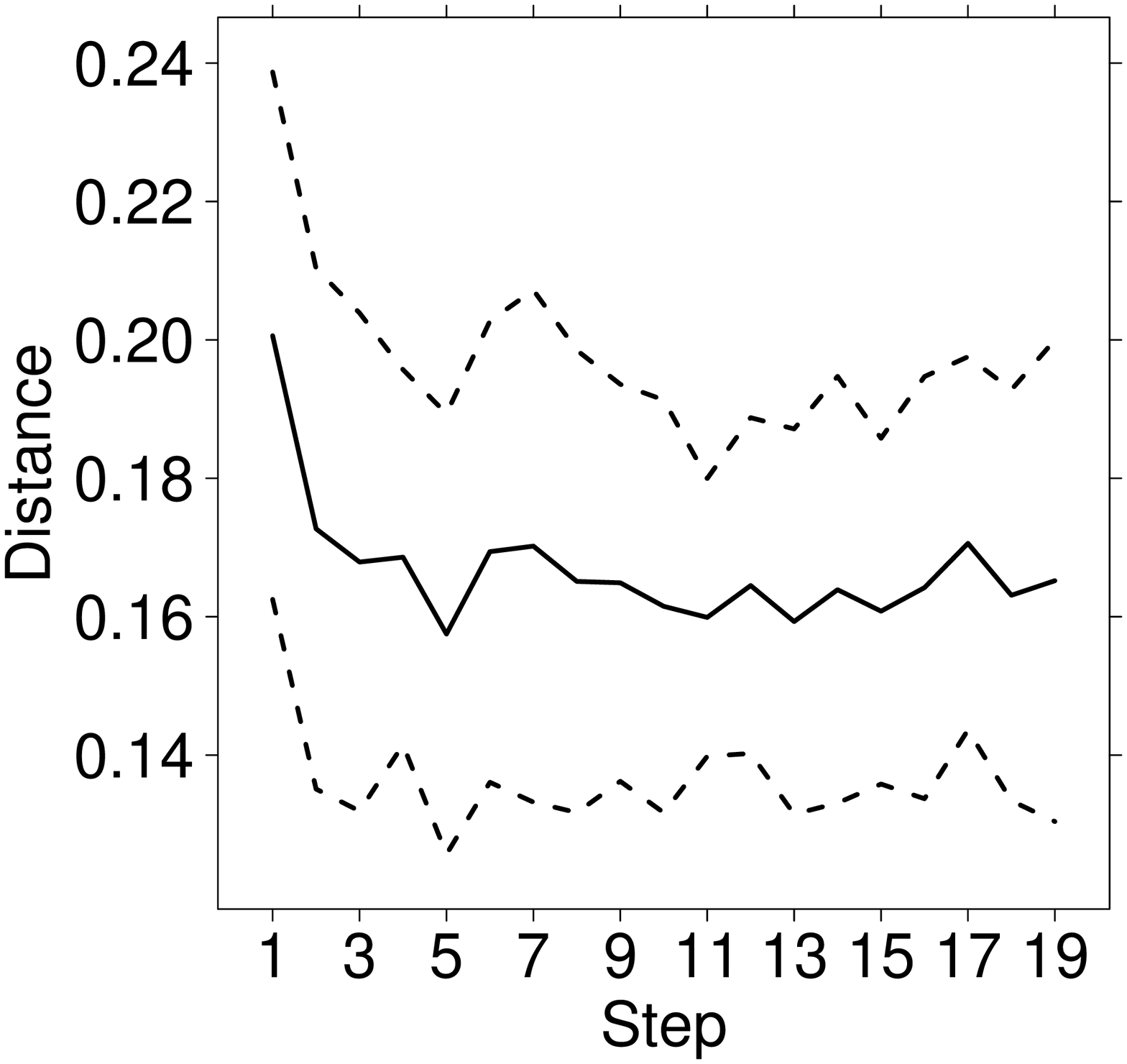}&
\includegraphics[width=3.2cm]{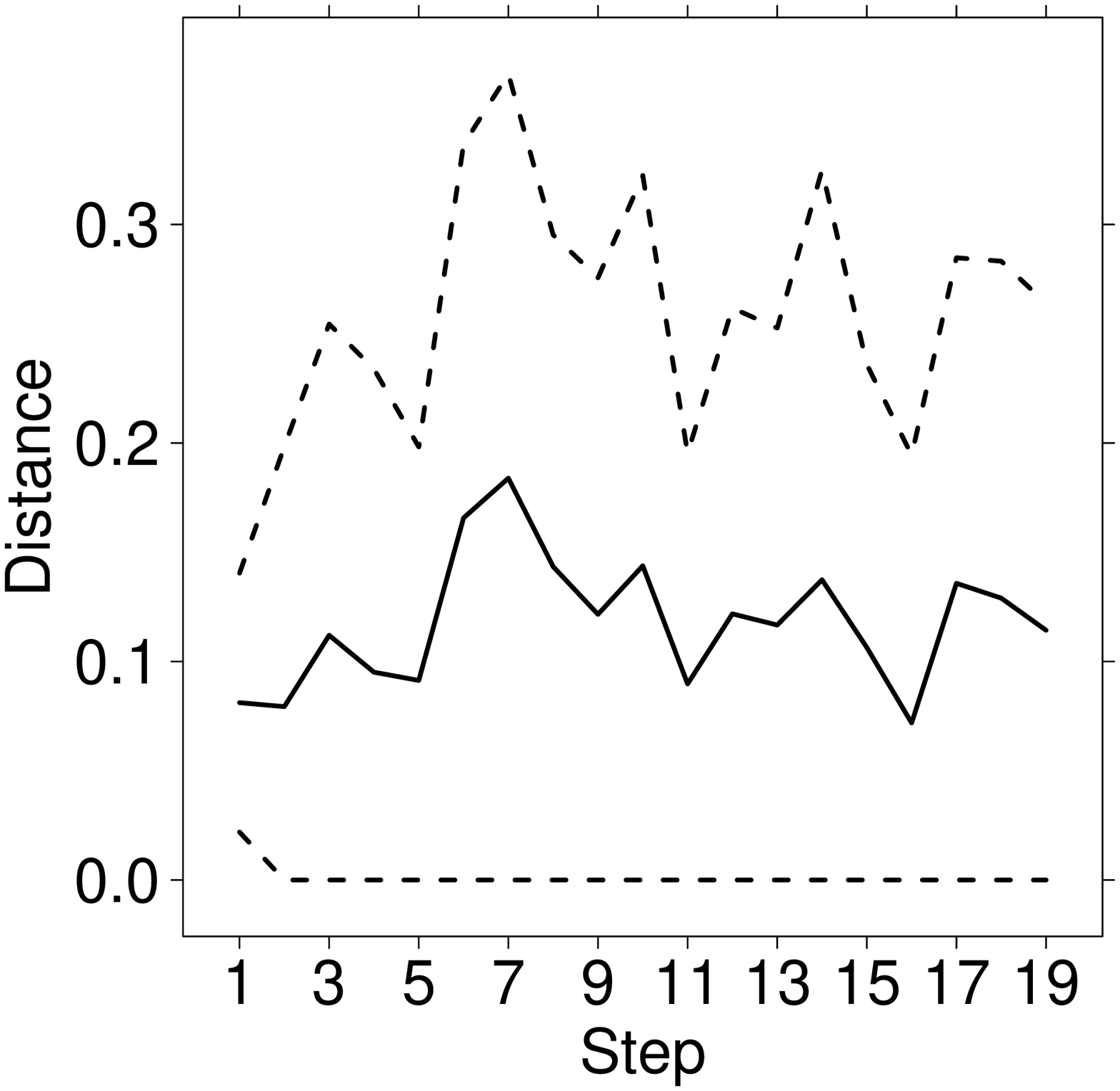}\\
D1 & D2 & D3\\
\includegraphics[width=3.2cm]{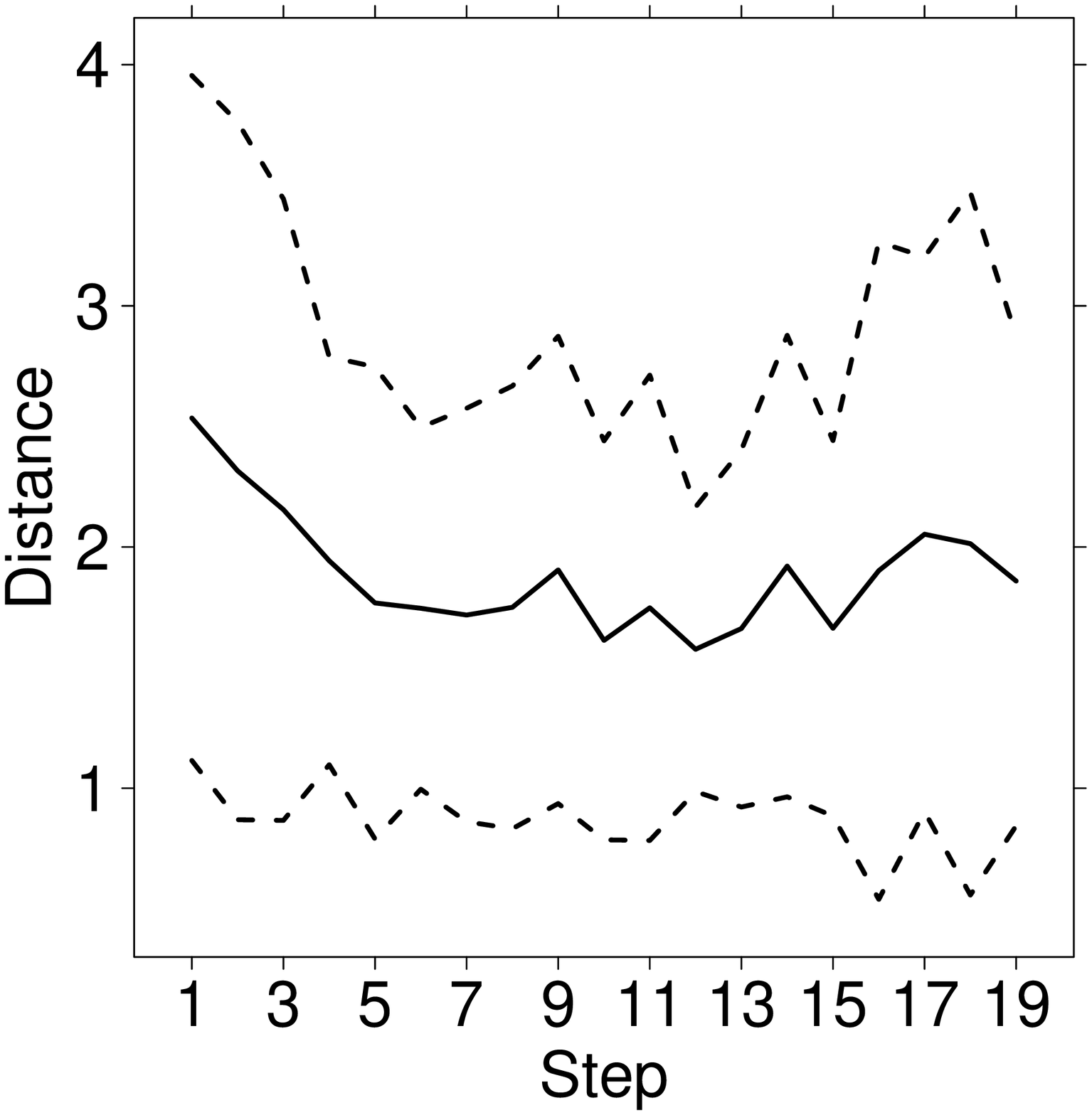}&
\raisebox{-0.2cm}{\includegraphics[width=3.6cm]{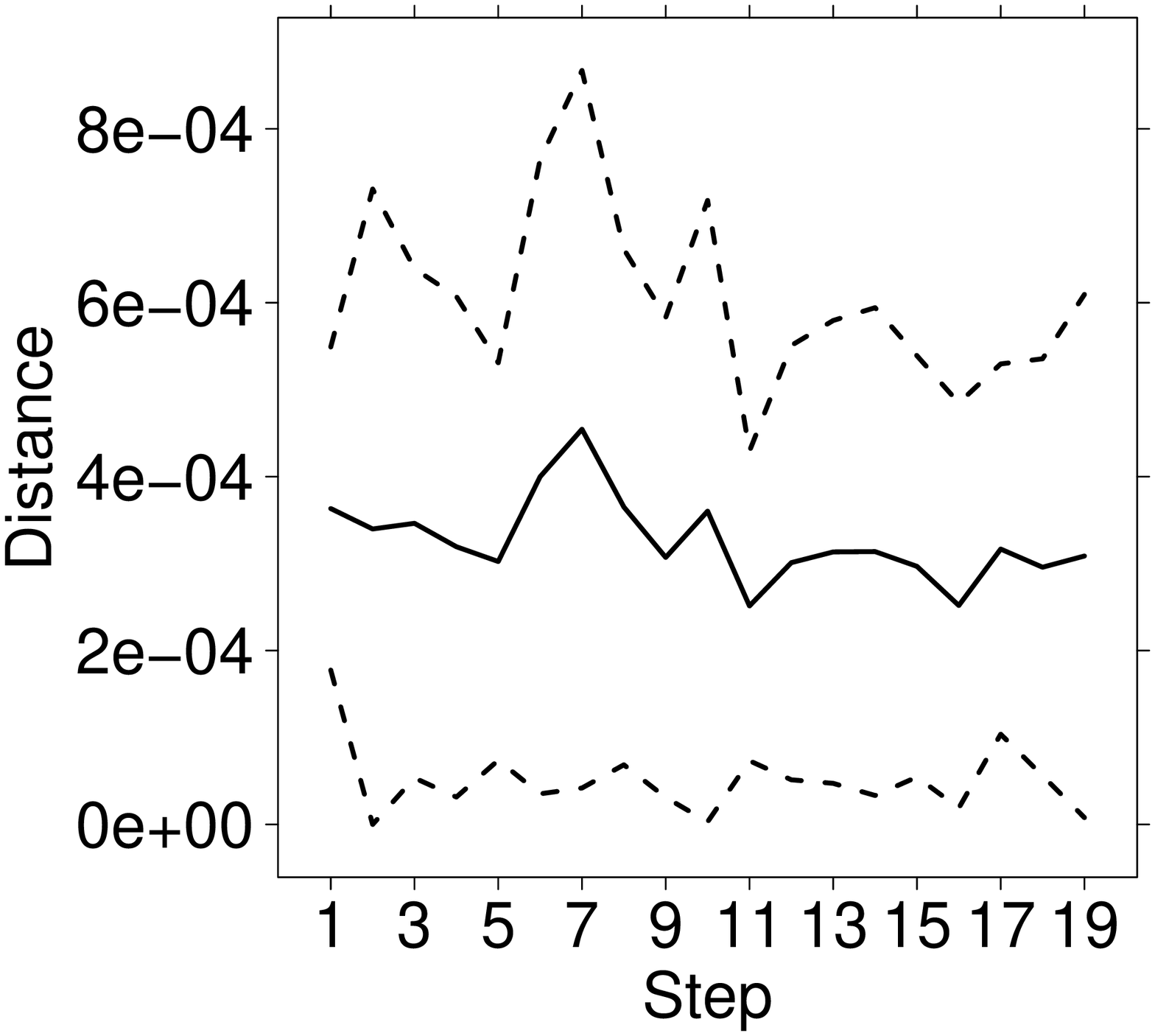}}&
\includegraphics[width=3.3cm]{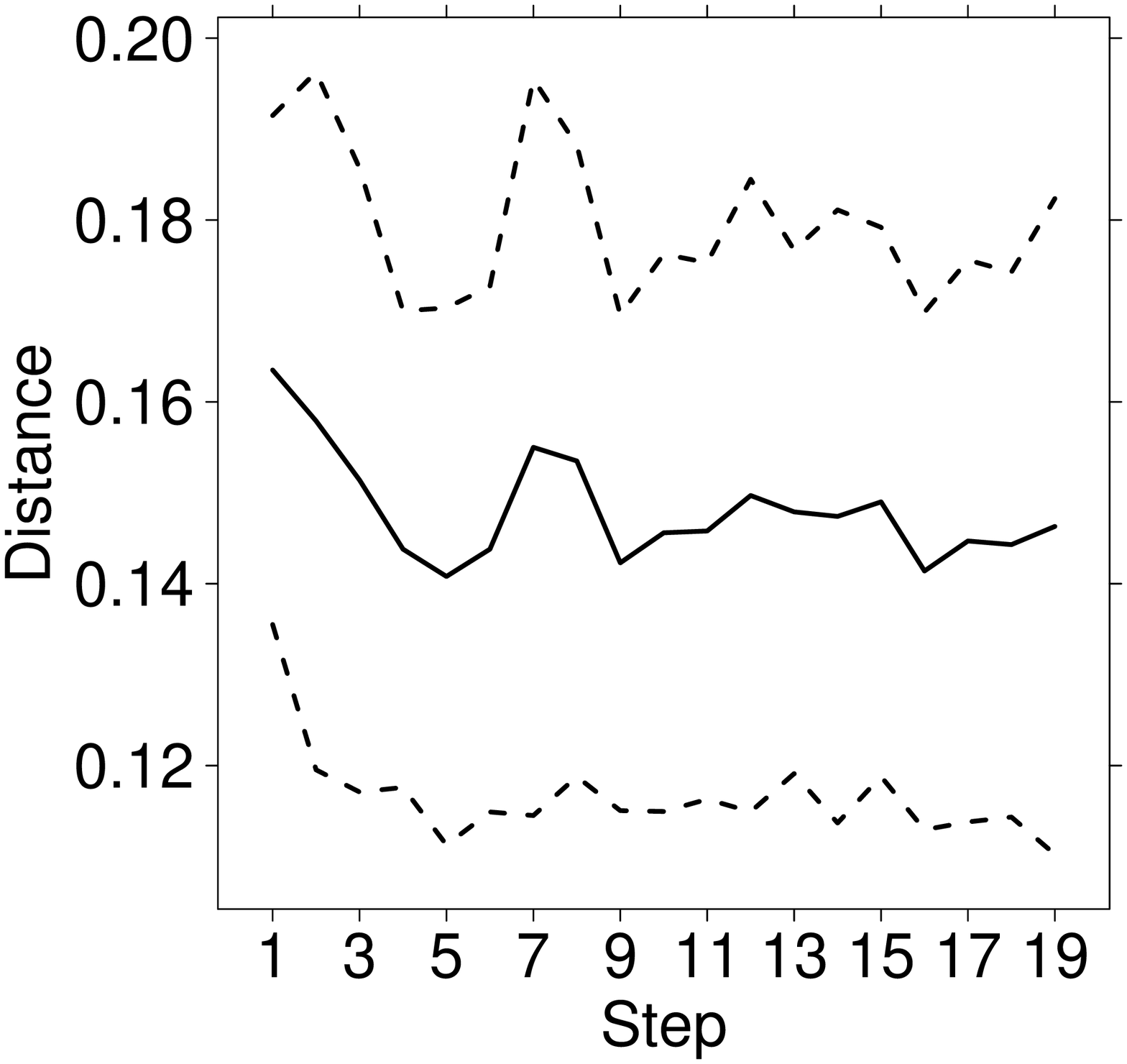}\\
D4 & D5 & D6 \\
\end{tabular}
}
\end{center}
\caption{Plots of the distances of consecutive elements of the series
of the dataset $\mathcal{B}_3(50,20,25,5,5)$. Solid line: mean
over the $b=50$ replicates; dashed lines: $1\sigma$ standard
deviation confidence intervals.}
\label{fig:exp3_trends}
\end{figure}

The Experiment 3 was performed on the benchmark dataset
$\mathcal{B}_3(50,20,25,5,5)$, and the results are reported in two
figures matching those of Exp. 2.  Since the difference between
consecutive pairs of elements of the series is quite similar
throughout all the steps, as expected all distances show a nearly
constant trend as shown in Fig. \ref{fig:exp3_trends}.

\begin{figure}[!ht]
\begin{center}
\includegraphics[width=0.75\textwidth]{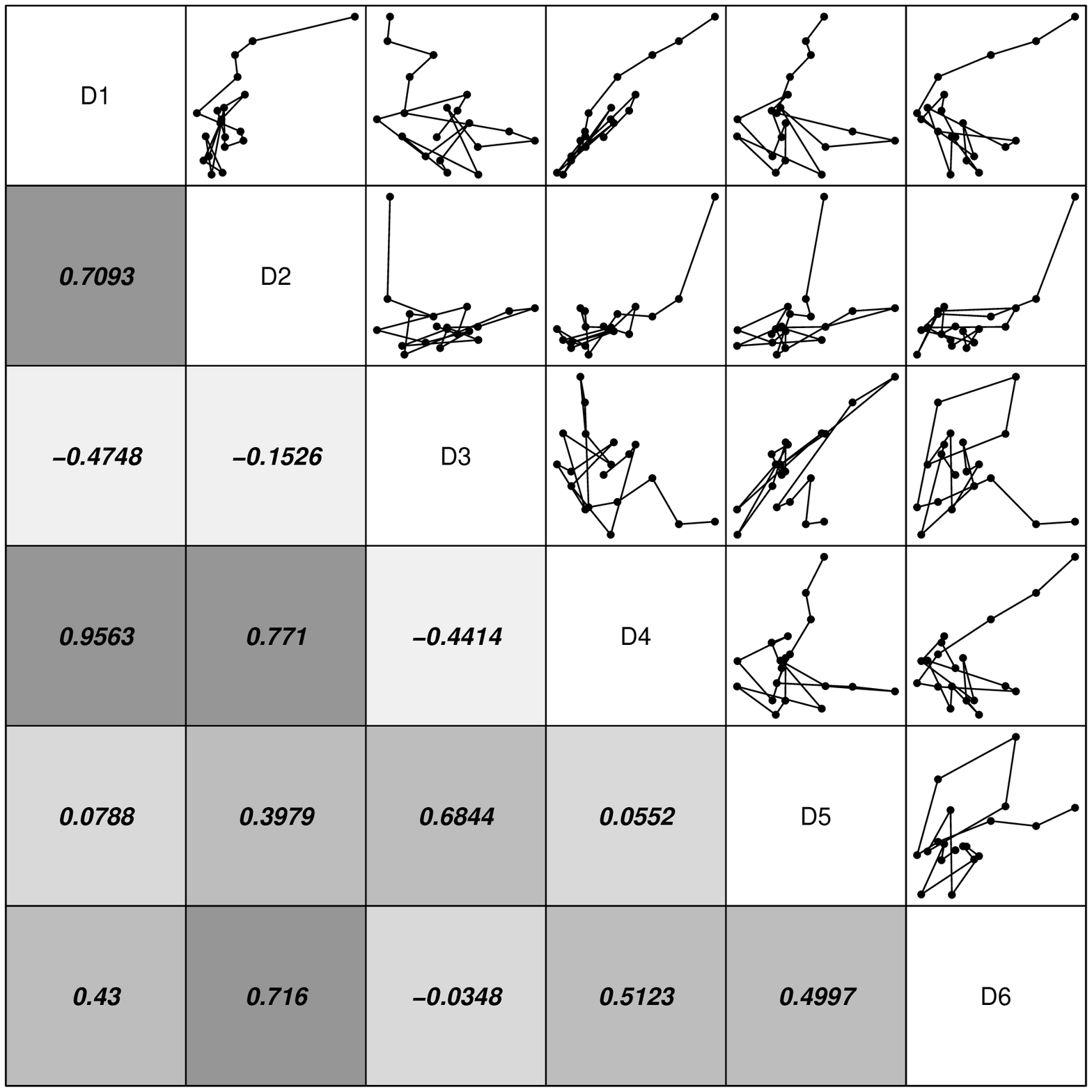}
\end{center}
\caption{Mutual scatterplots (upper triangle) and correlation values (lower triangle) for the Exp. 3.}
\label{fig:exp3_splom}
\end{figure}

The obscillations around the mean value are nevertheless strongly
varying among different measures, as evidenced by
Fig. \ref{fig:exp3_splom}. In particular, distance D3 is
anticorrelated to all distances but D5; furthermore only in 4 cases
out of 15 we obtain a correlation value higher than $0.7$, with again
$D1$, $D2$, $D4$ and $D6$ forming a group of more similar behaviour.

\begin{figure}[!ht]
\begin{center}
\begin{tabular}{cc}
\includegraphics[width=0.48\textwidth]{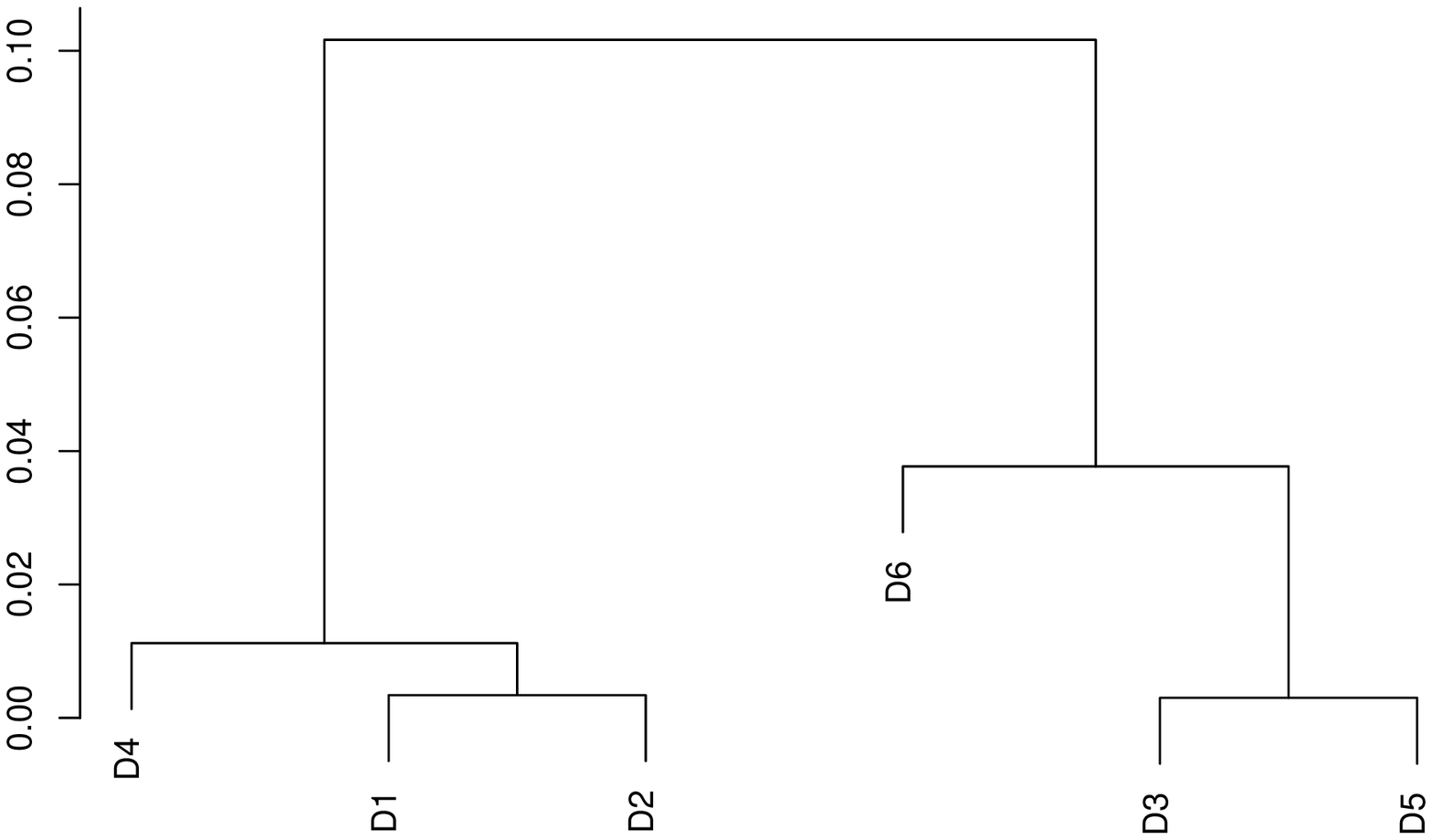} &
\includegraphics[width=0.48\textwidth]{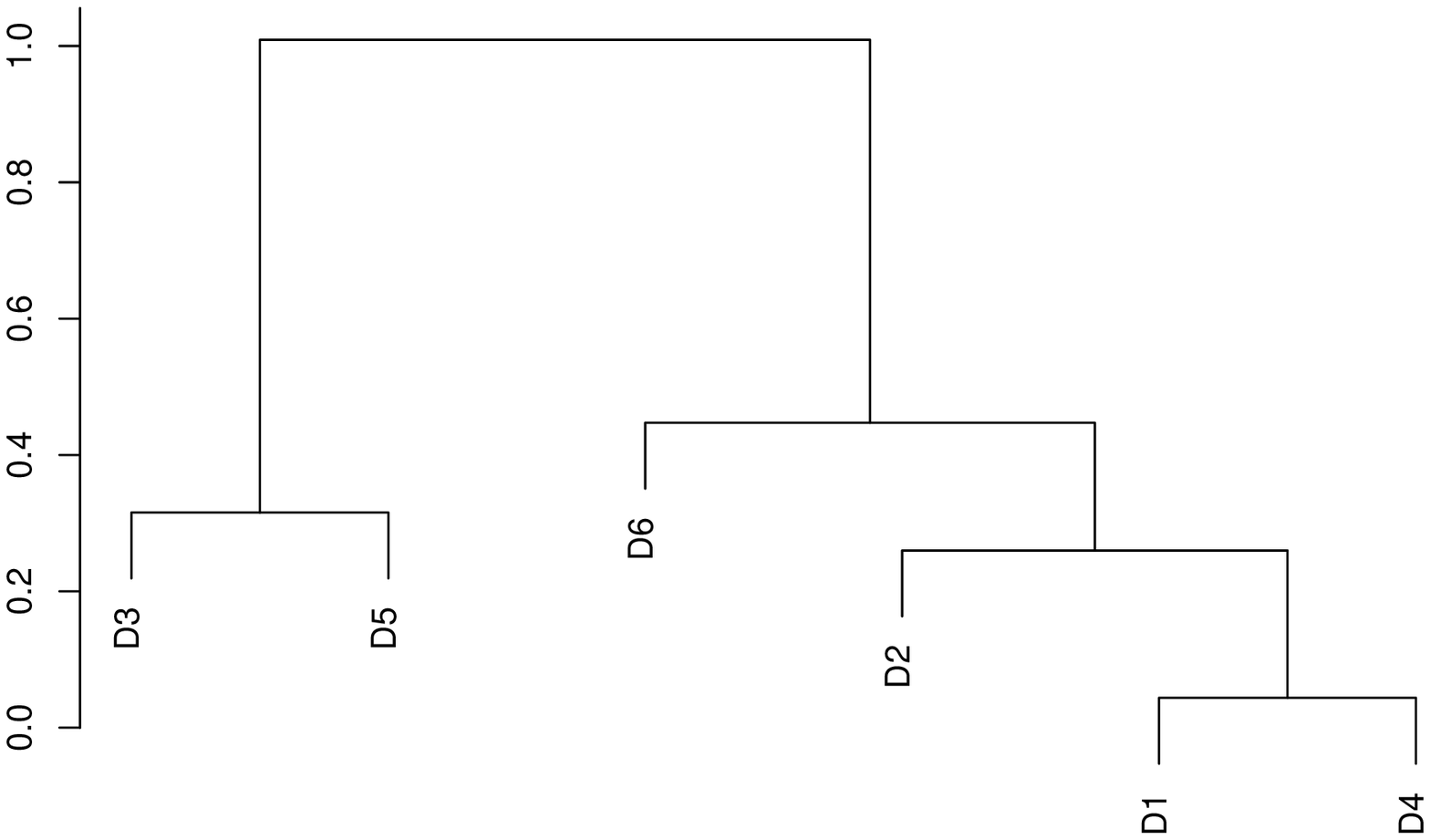}\\
Experiment 2 & Experiment 3\\
\end{tabular}
\end{center}
\caption{Cluster dendrograms with average linkage and correlation distance of D1-D6 for the two Experiments 2 and 3.}
\label{fig:dendrograms}
\end{figure}

Possible hierarchy of the six distances was explored by
clustering. Two dendrograms are built for Exp. 2 and Exp. 3 by using
the \textit{hclust} package in R and shown in
Fig. \ref{fig:dendrograms}. The clusters have average linkage and the
correlation distance
$\textrm{cd}(\cdot,\cdot)=1-\textrm{Corr}(\cdot,\cdot)$ is used as the
dissimilarity measure.  Although there is an appreciable coherence
among measures on macroscopic trends, when downscaling to microscopic
trends correlations get much looser. Distances $D1$, $D2$, $D4$, $D6$
seem to group together, while $D3$ has a more erratic
behaviour. Finally, a wide range difference occurs in the cluster
heights between the two experiments: the homogeneous macroscopic
situation of Exp. 2 has a narrower height span than the microscopic
case in Exp. 3.


\section{A regulatory network example}
\label{sec:grn}
\begin{figure}[!ht]                                                                                                                                      
\begin{center}   
\includegraphics[width=0.9\textwidth]{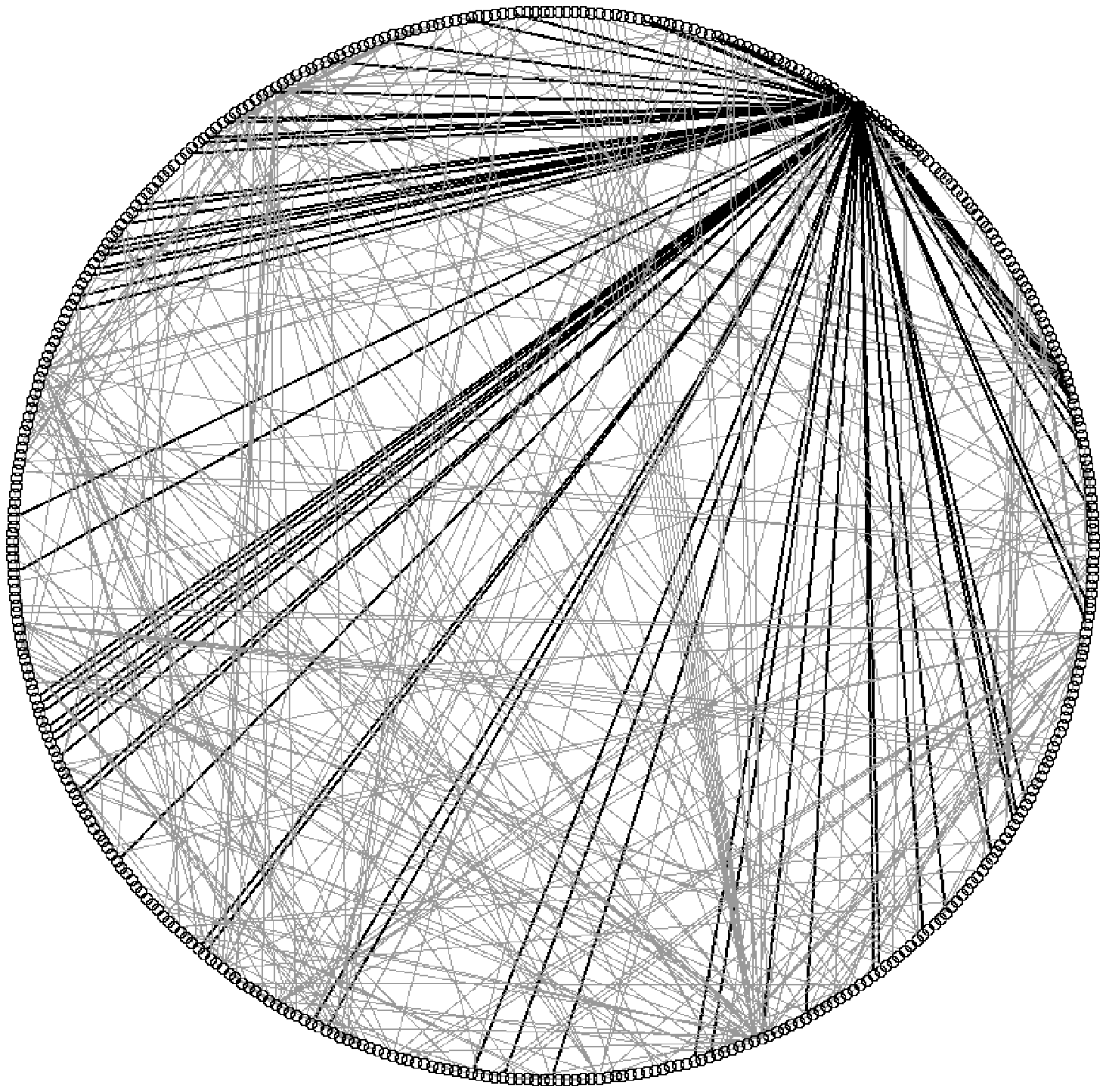} 
\end{center}
\caption{Transcriptional interactions network for \textit{Escherichia coli}, with edges relative to gene \textit{crp} marked in black.}
\label{fig:ecoli}
\end{figure}

To conclude with, we apply D1-D6 to three different perturbations of
the transcriptional interactions network\footnote{Publicly available
at
\url{http://www.weizmann.ac.il/mcb/UriAlon/Network_motifs_in_coli/ColiNet-1.1/}}
in \textit{Escherichia coli}, described in \cite{shen-orr02network} and shown
in Fig. \ref{fig:ecoli}.

The transcriptional database contains 577 interactions between 116 TFs
and 419 operons. Starting from an existing database
(RegulonDB\footnote{\url{http://regulondb.ccg.unam.mx/}}), the authors
added 35 new TFs, including alternative sigma factors, and over a
hundred new interactions from the literature.  The original adjacency
network (without self-interactions) consists of 420 vertices and 519
(undirected) links.  To show the influence on distances, we compare
the distances between the original network and the three networks
obtained by silencing out (thus deleting the link involving such
vertex) the activator/repressor factor \textit{crp} and the two
repressor factors \textit{frn} and \textit{himA}, having respectively
72, 22 and 21 links.  In Tab. \ref{tab:ecoli} we list the value of the
distances between the original network $EC$ and its three
perturbations, denoted respectively as $EC_{\overline{crp}}$,
$EC_{\overline{fnr}}$ and $EC_{\overline{himA}}$.

\begin{table}[h!]
\caption{Distances between $EC$ and the perturbed networks $EC_{\overline{crp}}$, $EC_{\overline{fnr}}$ and $EC_{\overline{himA}}$.}
\label{tab:ecoli}
\begin{center}
\begin{tabular}{c|c|rrrrrr}
\hline
\textbf{Network} & \textbf{Links} & \textbf{D1} & \textbf{D2} & \textbf{D3} & \textbf{D4} & \textbf{D5} & \textbf{D6}\\
\hline
($EC$,$EC_{\overline{crp}}$)                    & 519 vs 453 &  0.418 & 0.085 &  8.711  &  2191.9 &  1.01178$\cdot10^{-3}$ & 0.555 \\
($EC$,$EC_{\overline{fnr}}$)                    & 519 vs 497 &  0.058 & 0.023 &  0.191  &    41.4 &  0.01256$\cdot10^{-3}$ & 0.083 \\
($EC$,$EC_{\overline{himA}}$)                   & 519 vs 498 &  0.056 & 0.065 &  5.187  &    44.4 &  0.43315$\cdot10^{-3}$ & 0.404 \\
($EC_{\overline{crp}}$,$EC_{\overline{fnr}}$)   & 453 vs 497 &  0.557 & 0.074 &  6.938  &  2140.9 &  0.82079$\cdot10^{-3}$ & 0.479 \\
($EC_{\overline{crp}}$,$EC_{\overline{himA}}$)  & 453 vs 498 &  0.557 & 0.072 &  0.982  &  2138.1 &  0.26794$\cdot10^{-3}$ & 0.180 \\
($EC_{\overline{fnr}}$,$EC_{\overline{himA}}$)  & 497 vs 498 &  0.023 & 0.071 &  3.730  &    10.2 &  0.30303$\cdot10^{-3}$ & 0.357 \\
\hline
\end{tabular}
\end{center}
\end{table}

All distances seem to be heavily dependent on the number of removed
links: for all six distances, 
\begin{displaymath} 
D(EC,EC_{\overline{crp}}) > D(EC,EC_{\overline{fnr}}), D(EC,EC_{\overline{himA}})\ .
\end{displaymath}
Nevertheless, when the number of removed links are almost equal, such relation is
not valid anymore.

The distance $D(EC,EC_{\overline{himA}})$ is comparable to $D(EC,EC_{\overline{fnr}})$ for $D=D1, D4$, while the former is much bigger than the latter for all other distances. In fact, 
\begin{displaymath}
\frac{D(EC,EC_{\overline{himA}})}{D(EC,EC_{\overline{fnr}})} \; \simeq \; 
\begin{cases}
\;2.8 & \textrm{for D2}\\
\;4.8 & \textrm{for D6}\\
\\
\;27 & \textrm{for D3}\\
\;35 & \textrm{for D5}
\end{cases}
\end{displaymath}
For instance, the corresponding ratios $D(EC,EC_{\overline{crp}})/D(EC,EC_{\overline{himA}})$ are much smaller, namely
\begin{displaymath} 
\frac{D6(EC,EC_{\overline{crp}})}{D6(EC,EC_{\overline{himA}}}\simeq 1.4
\quad\textrm{and}\quad
\frac{D3(EC,EC_{\overline{crp}})}{D3(EC,EC_{\overline{himA}}}\simeq 1.7\ .
\end{displaymath}
  
A possible explanation is in the quite different structure of the two networks $EC_{\overline{fnr}}$ and $EC_{\overline{himA}}$, although being obtained silencing out almost the same number of links from the original network.

\begin{table}[!t]
\caption{Natural logarithm of the size of the automorphism group of the original and the perturbed networks}
\label{tab:auto}
\begin{center}
\begin{tabular}{c|rr}
\hline
\textbf{Network G} & $\mathbf{\log(|\textrm{Aut}(G)|)}$& $\max \{\lambda_i\}$ \\
\hline
$EC$                    & 330.0173 & 73.021\\
$EC_{\overline{crp}}$   & 377.5827 & 27.015\\
$EC_{\overline{fnr}}$   & 341.4692 & 73.019\\
$EC_{\overline{himA}}$  & 347.4488 & 73.020\\
\hline
\end{tabular}
\end{center}
\end{table}

The intrinsic structural difference between $EC_{\overline{fnr}}$ and $EC_{\overline{himA}}$ is indeed highlighted by the remarkable variation in the size of the respective group of automorphisms as shown in Tab. \ref{tab:auto}. For instance, the structure of $EC_{\overline{himA}}$ is almost $e^{347.4488-341.4692}\simeq 400$ times more symmetric than $EC_{\overline{fnr}}$. From this point of view, spectral distances can greatly help in analyzing subtle differences between networks where more classical methods are not helping much. As an example, the leading Laplacian eigenvalue is commonly used when network structure, because it is a good indicator of the stability and the local dynamics \cite{steuer08global}.  For instance, in this particular example, this value is of no help, as indicated in Tab. \ref{tab:auto}, since $EC$, $EC_{\overline{fnr}}$ and $EC_{\overline{himA}}$ have essentially the same leading eigenvalue; nevertheless the spectral distances, encoding information coming from the whole spectrum, can better separate very similar networks. Summarizing the observations following the experiments on synthetic data and the results in Tab. \ref{tab:ecoli}, we can conclude proposing $D2$ as the more reliable metric, both in terms of stability and robustness in terms of being less prone to odd behaviours.

\section*{Acknowledgements}
The authors acknowledge funding by the European Union FP7 Project HiperDART and by the Italian Ministry of Health Project ISITAD (RF 2007 conv. 42).
The authors are grateful to Samantha Riccadonna for her help with the R programming language.

\bibliographystyle{unsrt}
\bibliography{jurman10introduction}
\end{document}